\documentstyle[aps,epsf]{revtex}

\begin{document}

\title{Unified Classical and Quantum Radiation Mechanism for Ultra-Relativistic Electrons in Curved and Inhomogeneous
Magnetic Fields}
\author{T. Harko\footnote{E-mail: tcharko@hkusua.hku.hk} and K. S. Cheng\footnote{E-mail:hrspksc@hkucc.hku.hk}}
\address{Department of Physics, The University of Hong Kong,
Pokfulam Road, Hong Kong, P. R. China.}
\date{February 21, 2002}

\maketitle

\begin{abstract}

We analyze the general radiation emission mechanism from a charged particle moving in
a curved inhomogeneous magnetic field. The consideration of the gradient makes the curved
vacuum magnetic field compatible with the Maxwell equations and adds a non-trivial term to the
transverse drift velocity and, consequently, to the general radiation spectrum.
To obtain the radiation spectrum in the classical domain a general expression
for the spectral distribution and characteristic frequency of an electron in arbitrary
motion is derived by using
Schwinger's  method. The radiation patterns of the ultrarelativistic electron are represented 
in terms of the acceleration of the particle.
The same results can be obtained by considering that
the motion of the electron can be formally described as an evolution due to magnetic and electric forces.
By defining an effective electromagnetic field, which combines the magnetic field with the fictitious electric
field associated to the curvature and drift motion, one can obtain all the physical characteristics
of the radiation by replacing the constant magnetic field with the effective field.
The power, angular distribution and spectral distribution of all three components (synchrotron, curvature and gradient)
of the radiation are considered in both classical and quantum domain in the
framework of this unified formalism.
In the quantum domain the proposed approach allows the study of the effects of the
inhomogeneities and curvature of the magnetic field  on the radiative transitions rates of
electrons between low-lying Landau levels and the ground state.  
\end{abstract}

\keywords{radiation mechanism - polarization: magnetic fields: nonthermal-relativity: radiation transition rates - Landau levels}

\section{Introduction}

Since Ginzburg (1953), Shklovski (1953) and Gordon (1954) suggested that the
optical radiation of Crab Nebula was synchrotron radiation and would be
found to be polarized, the radiation of ultra-relativistic electrons    moving in magnetic fields had become a basic mechanism for the
understanding of radiation emission of astrophysical objects. At high
electron velocities the distribution of the radiation becomes a narrow cone
of half-angle $\theta $ given by $\theta =\gamma ^{-1}=\left( 1-\beta
^{2}\right) ^{1/2}=mc^{2}E^{-1}$ (Landau \& Lifshitz 1975), with $\beta =\frac{v}{c}$ and $E$ the
total energy of the electron. This directed beam of radiation is typical
regardless of the relation between the acceleration and velocity vectors.
The radiation is primarily directed in the direction transverse to the
magnetic field.

Synchrotron radiation plays a particularly important role in the
high energy emission of pulsars (Cheng, Ho \& Ruderman 1986a, Cheng, Ho \& Ruderman 1986b).
But in the radio region the curvature radiation,
resulting from the motion of particles along curved magnetic lines, plays an
equally important role (Ruderman \& Sutherland 1975). These two types of
radiation have been usually considered as separate mechanisms. A unifying
treatment, including in a single formalism the effects of both radiation
mechanisms, has been proposed by Zhang \& Cheng (1995) and Cheng \& Zhang (1996), called the
synchrocurvature radiation mechanism. In their approach the critical frequency of the
radiation of the electron is
\begin{equation}
\omega _{c}=\frac{3}{2}\gamma ^{3}\frac{c}{\rho }\left[ \frac{r_{B}^{2}+\rho
r_{B}-3\rho }{\rho r_{B}}\cos ^{4}\alpha +\frac{3\rho }{r_{B}}%
\cos ^{2}\alpha +\frac{\rho ^{2}}{r_{B}^{2}}\sin ^{4}\alpha \right] ^{1/2},
\end{equation}
where $\omega _{B}=\frac{eB}{\gamma mc}$, $r_{B}=\frac{c\sin \alpha }{%
\omega _{B}}$, $\alpha $ is the pitch angle and $\rho $ is the radius of curvature of the
local magnetic field. The total radiation power spectrum is given in the unified mechanism by
\begin{equation}
\frac{dI}{d\omega }=-\frac{\sqrt{3}e^{2}\gamma }{4\pi r_{c}^{\ast }}%
\left\{ \left[ F\left( \frac{\omega }{\omega _{c}}\right) -\frac{\omega }{%
\omega _{c}}K_{2/3}\left( \frac{\omega }{\omega _{c}}\right) \right] +\frac{%
\left[ \left( r_{B}+\rho \right) \Omega _{0}^{2}+r_{B}\omega _{B}^{2}\right]
^{2}}{c^{4}Q_{2}^{2}}\left[ F\left( \frac{\omega }{\omega _{c}}\right) +%
\frac{\omega }{\omega _{c}}K_{2/3}\left( \frac{\omega }{\omega _{c}}\right) %
\right] \right\}, 
\end{equation}
where $F(x)=x\int_{x}^{\infty }K_{5/3}(y)dy$, $\Omega _{0}=\frac{v\cos
\alpha }{\rho }$, $r_{c}^{\ast }=\frac{c^{2}}{\left( r_{B}+\rho \right)
\Omega _{0}^{2}+r_{B}\omega _{B}^{2}}$ and
\begin{equation}
Q_{2}^{2}=\frac{1}{r_{B}}\left(
\frac{r_{B}^{2}+\rho r_{B}-3\rho ^{2}}{\rho ^{3}}\cos ^{4}\alpha +\frac{3}{\rho }\cos
^{2}\alpha +\frac{1}{r_{B}}\sin ^{4}\alpha \right).
\end{equation}
$K_{n}$ denotes the MacDonald function of order $n$.

Due to the complicated character of the particle motion in a realistic physical
situation (the magnetic field of pulsars is of dipole form), to obtain the previous formulae the simplifying
assumption of a circular magnetic field with constant magnitude has been
assumed. This radiation mechanism generalizes all the classical results of
ordinary synchrotron and curvature radiation. Zhang \& Yuan (1998) have
studied the corresponding radiation mechanism in the quantum domain by
applying the method of equivalent photon scattering (Lieu \& Axford 1993; Lieu \& Axford 1995; Lieu, Axford \& McKenzie 1997). 
The curvature of the magnetic field could make electron's radiation spectrum
distinct from the power-law one, widely used in the astrophysical literatures (Zhang, Xia \& Yang 2000).

However, in obtaining the unified synchrocurvature mechanism (Zhang \& Cheng 1995; Cheng \& Zhang 1996), a
non-physical, curved constant magnitude magnetic field model has been used. Such a field does not satisfy the Maxwell field
equations in vacuum. On the other hand a curved physical magnetic field is
inhomogeneous, with the density of the magnetic field lines varying from
point to point. The inhomogeneity of the magnetic field creates a magnetic
gradient $\nabla B$, which generates a transverse drift motion of the
particle and gives a non-negligeable contribution to the radiation spectrum.

It is the purpose of the present paper to consider the general radiation
mechanism from a relativistic electron moving in an inhomogeneous curved magnetic field,
with the corrections due to the magnetic field gradient explicitly taken into
account. We call this mechanism generalized synchrocurvature radiation mechanism.
For this case the general motion of the electron can be analyzed by using the concept of
the guiding center (Alfven \& Falthammar 1963), which separates the motion $%
\vec{v}$ of a particle into motions $\vec{v}_{\mid \mid }$ parallel and $%
\vec{v}_{\perp }$ perpendicular to the field.

There are two basic methods for investigating the radiation spectrum from relativistic
charged particles. The first one is based on the explicit calculation of the
electromagnetic fields and the subsequent determination of the radiation characteristics
from the asymptotic form of the Poynting vector (Landau \& Lifshitz 1975; Jackson 1998). A
second method, without explicit reference to the fields, directly links the nature
of the source to the associated radiation (Schwinger, Tsai \& Erber 1976).
Both these methods are of considerable calculational complexity. In order
to obtain a general expression for the radiation spectrum of the electron moving
in the curved magnetic field we first derive, by using Schwinger's method, a general
expression for the classical radiation spectrum and characteristic frequency of the electron
in arbitrary motion. All the radiation characteristics of the ultra-relativistic particle
depend on the absolute value of the acceleration of the particle. With the use of these
general formulae the classical radiation spectrum of the electron in the inhomogeneous
magnetic field can immediately be obtained, once the acceleration of the particle is known.

The same results can also be obtained using some simple physical considerations. Since
the effect of the combined curvature and magnetic drift motion can be described in
terms of an effective electric field, the combination of this field with the magnetic field
gives an effective electromagnetic field. Substitution of this effective field in the
standard expressions for the synchrotron radiation in the presence of constant magnetic fields
leads to a consistent description of the generalized synchrocurvature radiation mechanism
in both classical and quantum domains.

In the case of an extremely strong magnetic field there are several effects which become important
in obtaining the synchrotron spectrum for high values of electron energy. If the peak in the
photon energy spectrum approaches the electron kinetic energy, then the electron tends to lose
almost all of their energy in single photon emission events and the discrete nature of the emission
cannot be ignored. The classical value for the critical radiation frequency, $\gamma ^2\omega _c$, exceeds the
electron kinetic energy when $B>\gamma (\gamma -1)B_{0}$, where $B_0=4.414\times 10^{13}G$
is the critical field strength (Akhiezer \& Shulga 1996).
Therefore the classical radiation formula can violate conservation of energy. Quantum effects will also be important
if the cyclotron energy approaches the electron rest mass, when even relativistic electrons
may occupy low Landau states. Quantum effects such as recoil of the emitting electrons
will affect the spectrum at low energies near the first few harmonics. Therefore it is important
to consider the supplementary quantum effects due to the magnetic field
inhomogeneities on the generalized synchrocurvature radiation mechanism and on
the radiative transitions rates for relativistic electrons in strong magnetic fields of the order
of $10^{12} G$. Such field strength lead to an extremely rapid energy losses of charged particles.

The present paper is organized as follows. The motion of electrons in curved
inhomogeneous magnetic fields is considered in Section 2. The corresponding
general radiation mechanism is described in the classical 
domain in Section 3. In Section 4 we discuss the quantum corrections to the
radiation spectrum. Radiation transitions from low Landau levels to the ground state are
discussed in Section 5. In Section 6 we discuss and conclude our results.

\section{Motion of charged particles in inhomogeneous curved magnetic fields}

Let us assume first a constant and homogeneous magnetic field $\vec{B}=B\vec{%
n}$, with $\vec{n}$ the unit vector directed along the magnetic field. The
equation of motion of an electron in this field is $\frac{d\vec{p}_{0}}{dt}=%
\frac{e}{c}\vec{v}_{0}\times \vec{B}$, where $\vec{p}_{0}=\frac{E}{c^{2}}%
\vec{v}_{0}$. $E$ and $\vec{v}_{0}$ are the total energy (which is a
constant of the motion) and the velocity of the particle, respectively. By
integrating the equation of motion we find $\vec{v}_{0}=\omega _{B}$ $\vec{r}%
\times \vec{n}+\vec{b}$, $\omega _{B}=\frac{eB}{\gamma mc}$ and $\vec{b}%
=const.$ (Jackson 1998). $\vec{r}$ is the position vector of the particle satisfying
the relation $\vec{n}\cdot \vec{r}=b_{\mid \mid }=v_{\mid \mid }(t=0)$.

Suppose that the magnetic field is homogeneous with the lines of force
curved with a radius $R_{c}$. In a vacuum such a field would obey the
Maxwell field equations only if combined with a magnetic field gradient. The
net drift therefore would be the sum of the curvature drift and the gradient
drift.

To derive the curvature drift alone, let us start from a simplified
configuration without a magnetic field gradient. Then the drift of the
particle arises from a centrifugal force $\vec{F}_{cf}$ acting on the
particle during its motion parallel to the field. Therefore the drift is
determined by the speed $\vec{v}_{\mid \mid }$ parallel to the field, where $%
\left| \vec{v}_{\mid \mid }\right| =v_{0}\cos \alpha $ and  $\alpha $ is the
angle between the vectors $\vec{v}_{0}$ and $\vec{B}$. The centrifugal force
gives rise to an acceleration of magnitude $\frac{v_{\mid \mid }}{R_{c}}$,
which can be viewed as arising from an effective electric field $\vec{E}%
_{eff}=\frac{\gamma m}{e}v_{\mid \mid }\frac{\vec{R}_{c}}{R_{c}^{2}}$
(Jackson 1998). This combined effective electric and magnetic fields cause a
curvature drift velocity $\vec{v}_{C}=c\frac{\gamma m}{e}v_{\mid \mid }^{2}%
\frac{\vec{R}_{c}\times \vec{B}}{R_{c}^{2}B^{2}}$.

In a real field, a field gradient would exist in addition to the curvature.
In a vacuum without electric currents the magnetic field obeys $\nabla
\times \vec{B}=0$. In cylindrical coordinates $\vec{B}$ has only one
component in $\theta $, $B_{\theta }\sim \frac{1}{r}$, $\nabla B$ has only
one in $r$ and $\nabla \times \vec{B}$ only one in $z$, given by $\left(
\nabla \times \vec{B}\right) _{z}=\frac{1}{r}\frac{\partial }{\partial r}%
\left( rB_{\theta }\right) =0$. Then $\left| \vec{B}\right| $ is
proportional to $\frac{1}{R_{c}}$ and $\frac{\nabla \left| \vec{B}\right| }{%
\left| \vec{B}\right| }=-\frac{\vec{R}_{c}}{R_{c}^{2}}$. Therefore the
gradient drift reads $\vec{v}_{\nabla B}=c\frac{\gamma m}{2e}v_{\perp }^{2}%
\frac{\vec{R}_{c}\times \vec{B}}{R_{c}^{2}B^{2}}$ (Jackson 1998), with $%
v_{\perp }=v_{0}\sin \alpha $ .

Combining all these results we obtain the velocity of a charged particle
moving in a curved inhomogeneous magnetic field in the form
\begin{equation}\label{1}
\vec{v}=\vec{v}_{0}+\frac{\gamma mc}{e}\frac{v_{\mid \mid }^{2}+\frac{1}{2}%
v_{\perp }^{2}}{R_{c}^{2}B^{2}}\vec{R}_{c}\times \vec{B}. 
\end{equation}

For a rigorous deduction of the drift velocity equation in curved
inhomogeneous magnetic fields see Somov(1994). The magnitude of the
velocity vector is given by
\begin{equation}\label{2}
v^{2}=\left| \vec{v}\right| ^{2}=v_{0}^{2}+\frac{2\gamma mcv_{0}}{e}\frac{%
v_{\mid \mid }^{2}+\frac{1}{2}v_{\perp }^{2}}{R_{c}B}\sin \alpha +\frac{%
\gamma ^{2}m^{2}c^{2}}{e^{2}}\left( \frac{v_{\mid \mid }^{2}+\frac{1}{2}%
v_{\perp }^{2}}{R_{c}B}\right) ^{2}, 
\end{equation}
where we have taken into account that $\vec{v}_{0}\cdot \left( \vec{R}%
_{c}\times \vec{B}\right) \neq 0$.

The acceleration of the particle follows from the equation of
motion $\frac{d\vec{v}}{dt}=\frac{e}{\gamma mc}\vec{v}\times \vec{B}$ and is
given by
\begin{equation}\label{3}
\vec{a}=\frac{d\vec{v}}{dt}=\frac{e}{\gamma mc}\vec{v}_{0}\times \vec{B}+%
\frac{v_{\mid \mid }^{2}+\frac{1}{2}v_{\perp }^{2}}{R_{c}^{2}}\vec{R}_{c}. 
\end{equation}

The magnitude of the acceleration is
\begin{equation}\label{4}
a^{2}=\left| \vec{a}\right| ^{2}=\frac{e^{2}}{\gamma ^{2}m^{2}c^{2}}%
v_{0}^{2}B^{2}\sin ^{2}\alpha +\frac{2ev_{0}B}{\gamma mc}\frac{v_{\mid \mid
}^{2}+\frac{1}{2}v_{\perp }^{2}}{R_{c}}\sin \alpha +\frac{\left( v_{\mid
\mid }^{2}+\frac{1}{2}v_{\perp }^{2}\right) ^{2}}{R_{c}^{2}}. 
\end{equation}

This equation can also be written in the form
\begin{equation}\label{5}
\left| \vec{a}\right| ^{2}=a_{0}^{2}\left( 1+\frac{\gamma mc}{ev_{0}B\sin
\alpha }\frac{v_{\mid \mid }^{2}+\frac{1}{2}v_{\perp }^{2}}{R_{c}}\right)
^{2}, 
\end{equation}
where $a_{0}=\frac{ev_{0}B\sin \alpha }{\gamma mc}$ is the acceleration of
the electron in the circular motion transversal to the magnetic field.

The acceleration and velocity vectors satisfy the important relation
$\vec{a}\cdot \vec{v}=0\Rightarrow \vec{a}\perp \vec{v}$,
which follows from $\vec{v}\cdot \vec{a}=\left( e/\gamma mc\right) \vec{v}%
\cdot \left( \vec{v}\times \vec{B}\right) =\left( e/\gamma mc\right) \vec{B}%
\cdot \left( \vec{v}\times \vec{v}\right) =0$.

On the other hand from the equation of motion of the electron given by Eq.(\ref{3})
we can see that an effective electric field $\vec{E}_{eff}$ can also be associated
to the combined curvature and magnetic gradient drift motion, with
\begin{equation}\label{7}
\vec{E}_{eff}=\frac{\gamma m}{e}\frac{v_{\mid \mid }^{2}+\frac{1}{2}v_{\perp
}^{2}}{R_{c}^{2}}\vec{R}_{c}. 
\end{equation}

This observation plays a fundamental role in obtaining the radiation
spectrum of the electron moving in curved inhomogeneous magnetic fields.

\section{Classical Radiation spectrum}

With the use of the previous results the basic characteristics of the
radiation of an ultra-relativistic electron moving along the field lines in
an inhomogeneous magnetic field can be easily calculated.

The total energy radiated per unit time $P=\frac{d\varepsilon }{dt}$ can be
obtained from the general expression (Landau and Lifshitz 1975) 
\begin{equation}  \label{8}
P=\frac{d\varepsilon }{dt}=\frac{2e^{2}}{3c^{3}}\frac{a^{2}-\frac{\left( 
\vec{v}\times \vec{a}\right) ^{2}}{c^{2}}}{\left( 1-\frac{v^{2}}{c^{2}}%
\right) ^{3}}=\frac{2e^{2}}{3c^{3}}\frac{a^{2}}{\left( 1-\frac{v^{2}}{c^{2}}%
\right) ^{2}},
\end{equation}
where we have taken into account that in the present model the acceleration
and velocity vectors are perpendicular. Then a simple calculation yields 
\begin{equation}  \label{a}
P=\frac{2e^{4}\gamma ^{2}}{3m^{2}c^{5}}\left[ v_{0}^{2}B^{2}\sin ^{2}\alpha +%
\frac{2mcv_{0}\gamma B}{e}\frac{v_{\mid \mid }^{2}+\frac{1}{2}v_{\perp }^{2}%
}{R_{c}}\sin \alpha +\frac{m^{2}c^{2}\gamma ^{2}}{e^{2}}\frac{\left( v_{\mid
\mid }^{2}+\frac{1}{2}v_{\perp }^{2}\right) ^{2}}{R_{c}^{2}}\right].
\end{equation}

In Eq. (\ref{a}) and in the followings we denote by $e$ the absolute value
of the electric charge of the electron, $e=\left| e\right| $.

In the limit $\alpha \rightarrow 0$ and $v_{0}\rightarrow c$ the particle is
moving along the magnetic field lines, and we also have $v_{\perp }=0$. Then
from Eq. (\ref{a}) we obtain 
\begin{equation}
P=\frac{2e^{2}}{3c^{3}}\gamma ^{4}\frac{v_{\mid \mid }^{4}}{R_{c}^{2}}.
\end{equation}

For $v_{\mid \mid }\rightarrow c$ we obtain the well-known formula of the
curvature radiation power: $P_{curv}=\frac{2e^{2}c\gamma ^{4}}{3R_{c}^{2}}$.

The limiting case of the synchrotron radiation emitted by an
ultra-relativistic ($v_{0}\approx c$) electron in a constant magnetic field
is recovered for $R_{c}\rightarrow \infty $, leading to 
\begin{equation}  \label{synch}
P_{synch}=\frac{2e^{4}B^{2}\sin ^{2}\alpha \gamma ^{2}}{3m^{2}c^{3}}.
\end{equation}

The variations of the power emitted by an ultra-relativistic electron moving
in a curved inhomogeneous magnetic field as a function of $\sin \alpha $, $%
\gamma $, the magnetic field $B$ and the radius of curvature $R_c$ are
represented in Figs. 1-4. The figures compare $P$ for the cases of the
synchrotron radiation ($R_{c}\rightarrow \infty $), curvature radiation ($%
\alpha \rightarrow 0$), the simple synchrocurvature effect and the general
synchrocurvature effect with the inhomogeneities of the magnetic field taken
into account.

\begin{figure}[h]
\epsfxsize=10cm \centerline{\epsffile{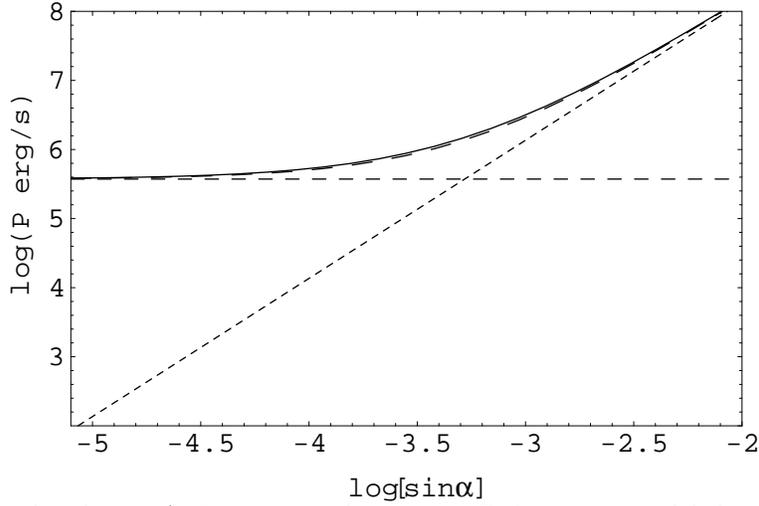}}
\caption{ Variation (in logarithmic scales) of the total emitted power $P$
of the ultrarelativistic electron as a function of $\sin \alpha$ for
the synchrotron emission (dotted curve), curvature radiation (short dashed
curve), simple synchrocurvature mechanism (Cheng \& Zhang 1996) (long dashed curve) and general
synchrocurvature model (solid curve) for $\gamma =3\times 10^7$, $%
R_c=10^8 cm$ and $B=10^6 G$. }
\label{FIG1}
\end{figure}

\begin{figure}[h]
\epsfxsize=10cm \centerline{\epsffile{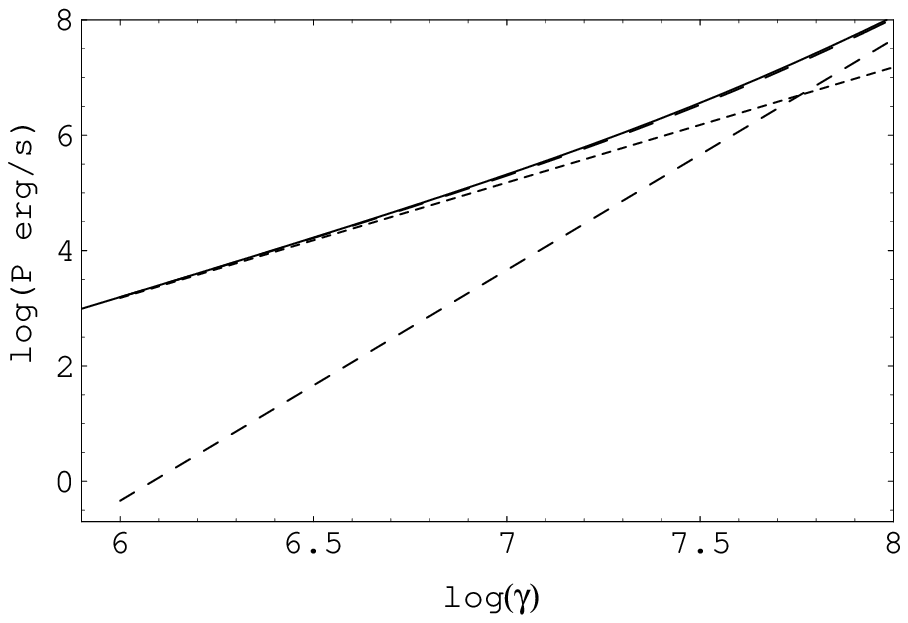}}
\caption{ Variation (in logarithmic scales) of the total emitted power $P$
of the ultrarelativistic electron as a function of $\gamma $ for the
synchrotron emission (dotted curve), curvature radiation (short dashed
curve), simple synchrocurvature mechanism (Cheng \& Zhang 1996) (long dashed curve) and general
synchrocurvature model (solid curve) for $\sin \alpha =10^{-3}$, $%
R_c=10^8 cm$ and $B=10^6 G$. }
\label{FIG2}
\end{figure}

\begin{figure}[h]
\epsfxsize=10cm \centerline{\epsffile{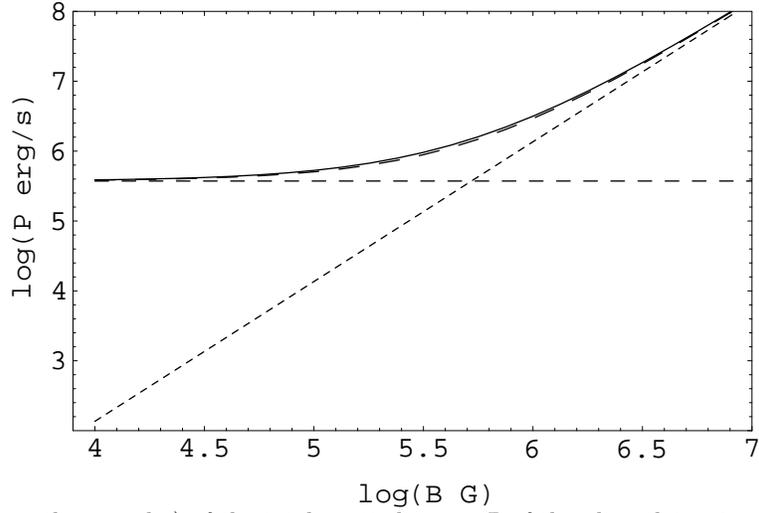}}
\caption{ Variation (in logarithmic scales) of the total emitted power $P$
of the ultrarelativistic electron as a function of the magnetic field $B$
for the synchrotron emission (dotted curve), curvature radiation (short
dashed curve), simple synchrocurvature mechanism (Cheng \& Zhang 1996) (long dashed curve) and
general synchrocurvature model (solid curve) for $\sin \alpha
=10^{-3}$, $R_c=10^8 cm$ and $\gamma =3\times 10^7$. }
\label{FIG3}
\end{figure}

\begin{figure}[h]
\epsfxsize=10cm \centerline{\epsffile{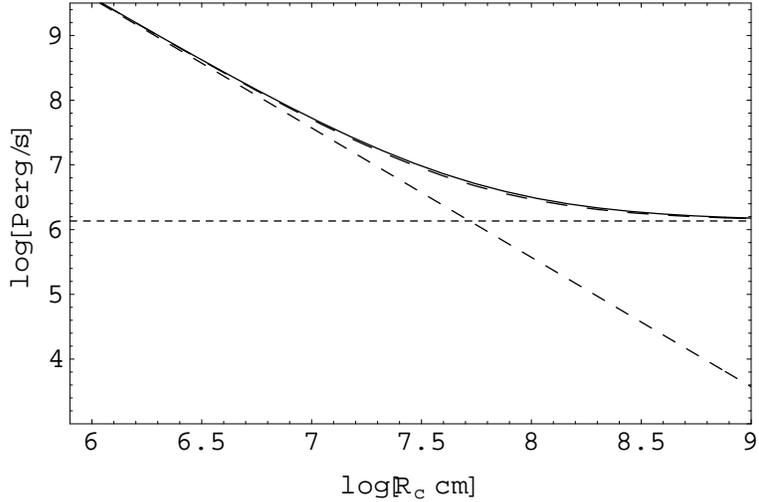}}
\caption{ Variation (in logarithmic scales) of the total emitted power $P$
of the ultrarelativistic electron as a function of the radius of curvature $%
R_c (cm)$ for the synchrotron emission (dotted curve), curvature radiation
(short dashed curve), simple synchrocurvature mechanism (Cheng \& Zhang 1996) (long dashed curve)
and general synchrocurvature model (solid curve) for $\sin \alpha
=10^{-3}$, $B=10^6 G$ and $\gamma =3\times 10^7$. }
\label{FIG4}
\end{figure}

We can see that the initial synchrocurvature results of (Zhang \& Cheng
1995; Cheng \& Zhang 1996) and the generalized synchrocurvature results
agree each other.

For the total radiation emitted into the solid angle $d\Omega $ we have
(Landau \& Lifshitz 1975) 
\begin{equation}  \label{9}
d\varepsilon _{\vec{n}}=\frac{e^{2}}{4\pi c^{3}}d\Omega \int \left[ \frac{%
2\left( \vec{n}\cdot \vec{a}\right) \left( \vec{v}\cdot \vec{a}\right) }{%
c\left( 1-\frac{\vec{v}\cdot \vec{n}}{c}\right) ^{4}}+\frac{\vec{a}^{2}}{%
\left( 1-\frac{\vec{v}\cdot \vec{n}}{c}\right) ^{3}}-\frac{\left( 1-\frac{%
v^{2}}{c^{2}}\right) \left( \vec{n}\cdot \vec{a}\right) ^{2}}{\left( 1-\frac{%
\vec{v}\cdot \vec{n}}{c}\right) ^{5}}\right] dt^{\prime },
\end{equation}
where $\vec{n}$ is the unit vector in the direction of the radiation and $%
t^{\prime }$ is the retarded time.

If the velocity and acceleration are perpendicular, for the intensity $dP$
radiated into the solid angle $d\Omega $ we find (Landau \& Lifshitz 1975) 
\begin{equation}
\frac{dP}{d\Omega }=\frac{e^{2}}{4\pi c^{3}}a^{2}\left[ \frac{1}{\left( 1-%
\frac{v}{c}\cos \theta \right) ^{4}}-\frac{\left( 1-\frac{v^{2}}{c^{2}}%
\right) \sin ^{2}\theta \cos ^{2}\phi }{\left( 1-\frac{v}{c}\cos \theta
\right) ^{6}}\right] ,  \label{10}
\end{equation}
where $\theta $ is the angle between $\vec{n}$ and $\vec{v}$ and $\phi $ is
the azimuthal angle of the vector $\vec{n}$ relative to the plane passing
through $\vec{v}$ and $\vec{a}$.

For the relativistic motion of the electron moving in inhomogeneous curved
magnetic fields we obtain 
\begin{eqnarray}
\frac{dP}{d\Omega } &=&\frac{e^{2}}{4\pi c^{3}}a_{0}^{2}\left( 1+\frac{%
\gamma mc}{ev_{0}B\sin \alpha }\frac{v_{\mid \mid }^{2}+\frac{1}{2}v_{\perp
}^{2}}{R_{c}}\right) ^{2}\times   \nonumber  \label{11} \\
&&\left[ \frac{1}{\left( 1-\frac{v}{c}\cos \theta \right) ^{4}}-\frac{\left(
1-\frac{v^{2}}{c^{2}}\right) \sin ^{2}\theta \cos ^{2}\phi }{\left( 1-\frac{v%
}{c}\cos \theta \right) ^{6}}\right] .
\end{eqnarray}

In order to find the spectral distribution of the classical radiation
emitted by an electron in the inhomogeneous curved magnetic field we shall
use the method introduced initially by Schwinger (1949) and further
developed in Schwinger, Tsai \& Erber (1976) (for an extensive presentation
of the method in the classical domain see also Schwinger, DeRaad, Milton \&
Tsai (1998)). By using this formalism we obtain first a general exact
expression for the radiation spectrum of an electric charge in arbitrary
motion. 

The total radiated power per unit frequency emitted by a relativistic electric charge in
motion on an arbitrary trajectory is given by the following expression (Schwinger 1949;
Schwinger, Tsai \& Erber 1976)
\begin{equation}
\frac{dP\left( \omega ,t\right) }{d\omega }=-\frac{\omega }{4\pi ^{2}}\int d%
\vec{r}d\vec{r}^{\prime }dt^{\prime }\cos \left[ \omega \left( t-t^{\prime
}\right) \right] \frac{\sin \left[ \frac{\omega }{c}\left| \vec{r}-\vec{r}%
^{\prime }\right| \right] }{\left| \vec{r}-\vec{r}^{\prime }\right| }\left[
\rho \left( \vec{r},t\right) \rho \left( \vec{r}^{\prime },t^{\prime
}\right) -\frac{1}{c^{2}}\vec{j}\left( \vec{r},t\right) \vec{j}\left( \vec{r}%
^{\prime },t^{\prime }\right) \right] ,  \label{sch1}
\end{equation}
where $\rho $ and $\vec{j}$ are the electric charge and current densities,
respectively. For a point electron of charge $e$, located at the variable
arbitrary position $\vec{R}(t)$, we have $\rho \left( \vec{r},t\right)
=e\delta \left( \vec{r}-\vec{R}(t)\right) $and $\vec{j}=e\vec{v}(t)\delta
\left( \vec{r}-\vec{R}(t)\right) $, where $\vec{v}(t)=\frac{d\vec{R}}{dt}$
is the velocity of the electron. With the use of these expressions for $\rho 
$ and $\vec{j}$ the spatial integrations in Eq. (\ref{sch1}) can be
performed and after some simple transformations we obtain for the radiation
spectrum the following representation:
\begin{equation}
\frac{dP\left( \omega ,t\right) }{d\omega }=-\frac{\omega e^{2}}{\pi }%
\int_{-\infty }^{+\infty }\left[ 1-\frac{\vec{v}(t)\cdot \vec{v}\left(
t+\tau \right) }{c^{2}}\right] \frac{\sin \left[ \frac{\omega }{c}\left| 
\vec{R}\left( t+\tau \right) -\vec{R}\left( t\right) \right| \right] }{%
\left| \vec{R}\left( t+\tau \right) -\vec{R}\left( t\right) \right| }\cos
\left( \omega \tau \right) d\tau .  \label{sch2}
\end{equation}

To evaluate the different terms in Eq. (\ref{sch2}) we use the power series 
representation of $\vec{R}\left( t+\tau \right) $ and $\vec{v}(t)$. Hence we
first obtain 
\begin{equation}
\left| \vec{R}\left( t+\tau \right) -\vec{R}\left( t\right) \right| =\left| 
\vec{v}\tau +\frac{\tau ^{2}}{2}\frac{d\vec{v}}{dt}+\frac{\tau ^{3}}{6}\frac{%
d^{2}\vec{v}}{dt^{2}}\right| \approx \left| \vec{v}\right| \tau -\frac{%
\left| \vec{a}\right| ^{2}}{c}\frac{\tau ^{3}}{24}.  \label{sch3}
\end{equation}

To obtain Eq. (\ref{sch3}) we have taken into account that $\vec{v}\cdot 
\frac{d\vec{v}}{dt}=\frac{1}{2}\frac{d\vec{v}^{2}}{dt}\approx 0$ in the
extrem limit of high velocities  $v\rightarrow c$ and $\vec{v}\cdot \frac{%
d^{2}\vec{v}}{dt^{2}}=\frac{d}{dt}\left( \vec{v}\cdot \frac{d\vec{v}}{dt}%
\right) -\left( \frac{d\vec{v}}{dt}\right) ^{2}\approx -\vec{a}^{2}$. In the
second term of Eq. (\ref{sch3}) we have also substituted $v$ by $c$.

By means of similar transformations we can evaluate the velocity dependent term
in the radiation spectrum:
\begin{equation}
1-\frac{\vec{v}(t)\cdot \vec{v}\left( t+\tau \right) }{c^{2}}=1-\beta ^{2}+%
\frac{\left| \vec{a}\right| ^{2}}{c^{2}}\frac{\tau ^{2}}{2}.  \label{sch4}
\end{equation}

Therefore the radiation spectrum can be represented in the form
\begin{equation}
\frac{dP\left( \omega ,t\right) }{d\omega }=-\frac{2\omega e^{2}}{\pi c}%
\int_{0}^{\infty }\left( 1-\beta ^{2}+\frac{\left| \vec{a}\right| ^{2}}{c^{2}%
}\frac{\tau ^{2}}{2}\right) \frac{\sin \left[ \omega \left( \beta \tau -%
\frac{\left| \vec{a}\right| ^{2}}{c^{2}}\frac{\tau ^{3}}{24}\right) \right] 
}{\tau }\cos \left( \omega \tau \right) d\tau .  \label{sch5}
\end{equation}

In the denominator of Eq. (\ref{sch5}) we have considered only the first
order approximation for $\left| \vec{R}\left( t+\tau \right) -\vec{R}\left(
t\right) \right| $ and the high velocity limit with $v\approx c$. By
transforming the product of the trigonometric functions in sum the above
integral becomes:
\begin{equation}
\frac{dP\left( \omega ,t\right) }{d\omega }=\frac{\omega e^{2}}{\pi c}%
\int_{0}^{\infty }\left( 1-\beta ^{2}+\frac{\left| \vec{a}\right| ^{2}}{c^{2}%
}\frac{\tau ^{2}}{2}\right) \left\{ \sin \left[ \omega \left( 1-\beta
\right) \tau +\omega \frac{\left| \vec{a}\right| ^{2}}{c^{2}}\frac{\tau ^{3}%
}{24}\right] -\sin \left( 2\omega \tau \right) \right\} \frac{d\tau }{\tau }.
\label{sch6}
\end{equation}

Introducing a new variable $\tau =2\frac{c}{\left| \vec{a}\right| }\sqrt{%
1-\beta ^{2}}x$, Eq. (\ref{sch6}) takes the form
\begin{equation}
\frac{dP\left( \omega ,t\right) }{d\omega }=\frac{\omega e^{2}}{\pi c}\left(
1-\beta ^{2}\right) \int_{0}^{\infty }\left( 1+2x^{2}\right) \left\{ \sin %
\left[ \frac{3}{2}\xi \left( x+\frac{1}{3}x^{3}\right) \right] -\sin
bx\right\} \frac{dx}{x},  \label{sch7}
\end{equation}
where $b=4\omega \frac{c}{\left| \vec{a}\right| }\sqrt{1-\beta ^{2}}$ and $%
\xi =\frac{2}{3}\omega \frac{c}{\left| \vec{a}\right| }\left( 1-\beta
^{2}\right) ^{3/2}$.

As is shown in Appendix, the above integral is equal
to $\frac{1}{\sqrt{3}}\int_{\xi }^{\infty }K_{5/3}\left( \eta \right) d\eta $.
Hence we obtain the following general form of the classical radiation spectrum of an
ultrarelativistic electron in arbitrary motion:
\begin{equation}
\frac{dP\left( \omega ,t\right) }{d\omega }=\frac{\sqrt{3}e^{2}}{2\pi }\frac{%
\left| \vec{a}\right| }{c^{2}}\gamma F\left( \frac{\omega }{\omega _{c}}%
\right) ,  \label{sch8}
\end{equation}
where the critical frequency $\omega _{c}$ is given by
\begin{equation}
\omega _{c}=\frac{3}{2}\frac{\left| \vec{a}\right| }{c}\gamma ^{3}.
\label{sch9}
\end{equation}

Therefore the radiation spectrum of an electron in arbitrary motion is
entirely determined by the magnitude of the acceleration of the particle along its trajectory.

In order to find the spectral distribution of the radiation emitted by an
electron in the inhomogeneous curved magnetic field we use the expression
for the acceleration of the electron, given by Eq. (\ref{4}), in Eqs. (\ref{sch8}) and (\ref{sch9}), thus obtaining:  
\begin{equation}
dP=d\omega \frac{\sqrt{3}}{2\pi }\frac{e^{3}}{mc^{2}}\sqrt{B^{2}\sin
^{2}\alpha +2\frac{\gamma mB}{e}\frac{v_{\mid \mid }^{2}+\frac{1}{2}v_{\perp
}^{2}}{R_{c}}\sin \alpha +\frac{\gamma ^{2}m^{2}}{e^{2}}\frac{\left( v_{\mid
\mid }^{2}+\frac{1}{2}v_{\perp }^{2}\right) ^{2}}{R_{c}^{2}}}F\left( \frac{%
\omega }{\omega _{c}}\right) ,  \label{p}
\end{equation}
where the critical frequency of the unified radiation spectrum is defined by 
\begin{equation}\label{17}
\omega _{c}=\frac{3e}{2mc}\sqrt{B^{2}\sin ^{2}\alpha +2\frac{\gamma mB}{e}%
\frac{v_{\mid \mid }^{2}+\frac{1}{2}v_{\perp }^{2}}{R_{c}}\sin \alpha +\frac{%
\gamma ^{2}m^{2}}{e^{2}}\frac{\left( v_{\mid \mid }^{2}+\frac{1}{2}v_{\perp
}^{2}\right) ^{2}}{R_{c}^{2}}}\left( \frac{E}{mc^{2}}\right) ^{2}.
\end{equation}

To obtain Eqs. (\ref{p}) and (\ref{17}) we have also considered the limit $%
v_{0}\rightarrow c$ in the expression of the acceleration of the particle
moving along the curved lines of the inhomogeneous magnetic field.

For $R_{c}\rightarrow \infty $, from Eq. (\ref{17}) we obtain the well-known
synchrotron critical frequency $\omega _{synch}=\frac{3eB\sin \alpha }{2mc}%
\left( \frac{E}{mc^{2}}\right) ^{2}$ (Landau \& Lifshitz 1975). In the limit $\alpha \rightarrow 0$,
the motion of the particle is along the magnetic field lines and the
radiation is the curvature radiation. By taking $v_{\perp }=0$ and $v_{\mid
\mid }\approx c$ , from Eq. (\ref{17}) we obtain $\omega _{curv}=\frac{3}{2}%
\left( \frac{E}{mc^{2}}\right) ^{3}\frac{c}{R_{c}}$. In the limit $%
R_{c}\rightarrow \infty $, from Eq.(\ref{p}) we obtain the 
spectral distribution of the synchrotron radiation, $dP=d\omega \frac{\sqrt{3%
}}{2\pi }\frac{e^{3}B\sin \alpha }{mc^{2}}F\left( \frac{\omega }{\omega _{c}}%
\right) $ (Landau \& Lifshitz 1975). 

In Figs. 5-8 we present the comparison of the behavior of the critical
frequency of the radiation, obtained from Eq. (\ref{17}) , with the
values obtained by using the synchrocurvature emission mechanism of (Cheng
\& Zhang 1995; Cheng \& Zhang 1996) and with the synchrotron and curvature
radiation mechanisms, respectively. The agreement between the generalized
synchrocurvature radiation in inhomogeneous magnetic fields and the
synchrocurvature mechanism of Cheng \& Zhang (1996) is very good.

\begin{figure}[h]
\epsfxsize=10cm \centerline{\epsffile{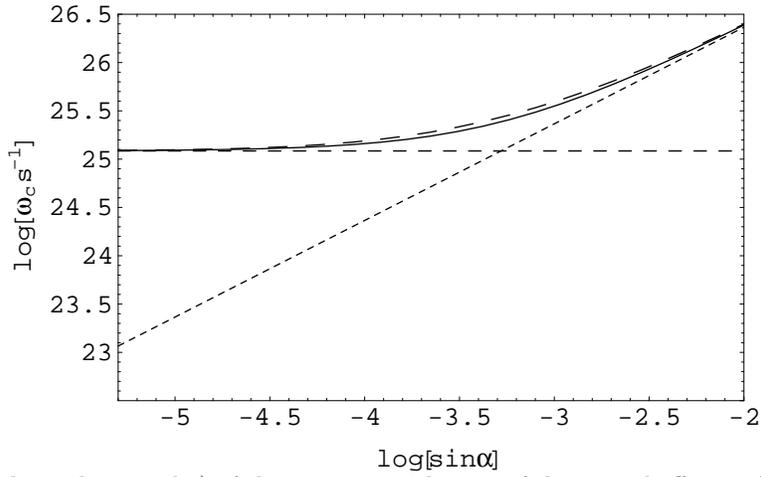}}
\caption{ Comparison (in logarithmic scales) of the variation with $\sin 
\alpha $ of the critical effective frequency in the generalized
synchrocurvature model (solid curve) with the critical frequency of the
synchrotron radiation (dotted curve), of the curvature radiation (short
dashed curve) and of the synchrocurvature mechanism proposed by Cheng \&
Zhang (1996) (long dashed curve) for $\gamma =3 \times 10^7$, $%
B=10^6 G$ and $R_c=10^8 cm$. }
\label{FIG5}
\end{figure}

\begin{figure}[h]
\epsfxsize=10cm \centerline{\epsffile{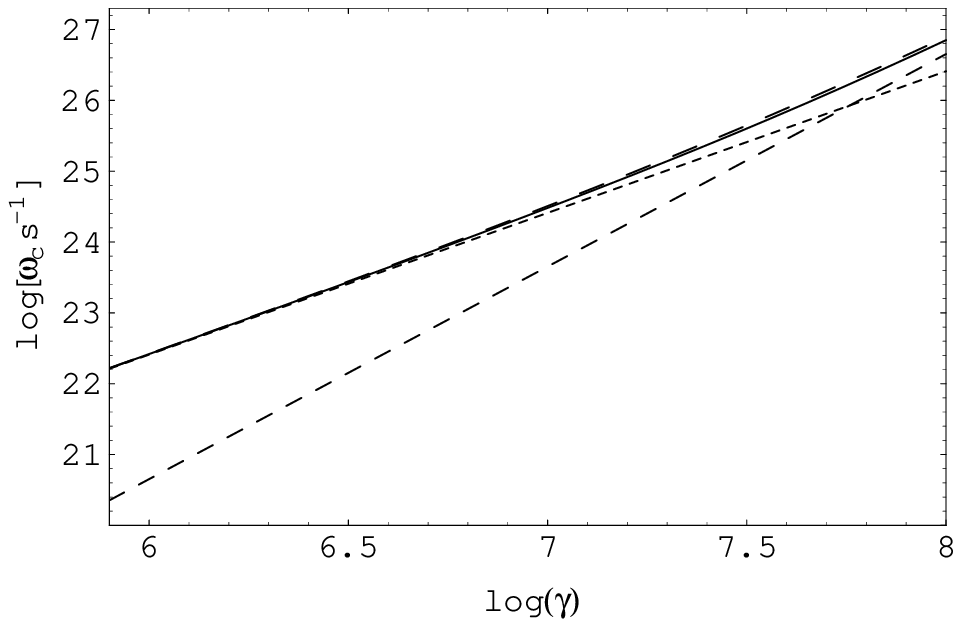}}
\caption{ Comparison (in logarithmic scales) of the variation with $\gamma $ of the critical effective frequency in the generalized
synchrocurvature model (solid curve) with the critical frequency of the
synchrotron radiation (dotted curve), of the curvature radiation (short
dashed curve) and of the synchrocurvature mechanism proposed by Cheng \&
Zhang (1996) (long dashed curve) for $\sin \alpha =10^{-3}$, $B=10^6
G$ and $R_c=10^8 cm$. }
\label{FIG6}
\end{figure}

\begin{figure}[h]
\epsfxsize=10cm \centerline{\epsffile{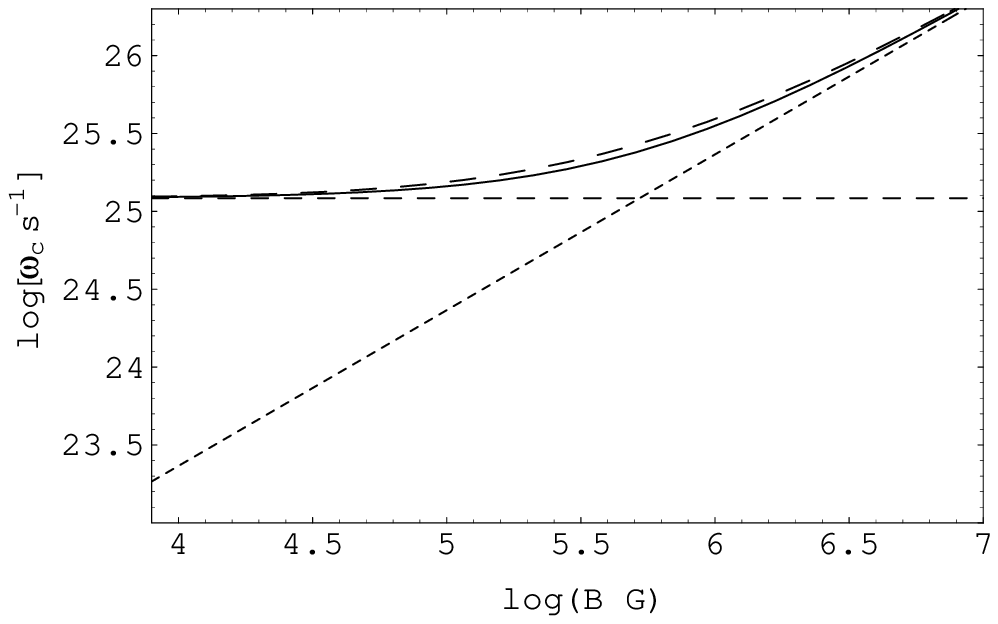}}
\caption{ Comparison (in logarithmic scales) of the variation with the
magnetic field $B$ of the critical effective frequency in the generalized
synchrocurvature model (solid curve) with the critical frequency of the
synchrotron radiation (dotted curve), of the curvature radiation (short
dashed curve) and of the synchrocurvature mechanism proposed by Cheng \&
Zhang (1996) (long dashed curve) for $\sin \alpha =10^{-3}$, $\gamma =3\times 10^7$ and $R_c=10^8 cm$. }
\label{FIG7}
\end{figure}

\begin{figure}[h]
\epsfxsize=10cm \centerline{\epsffile{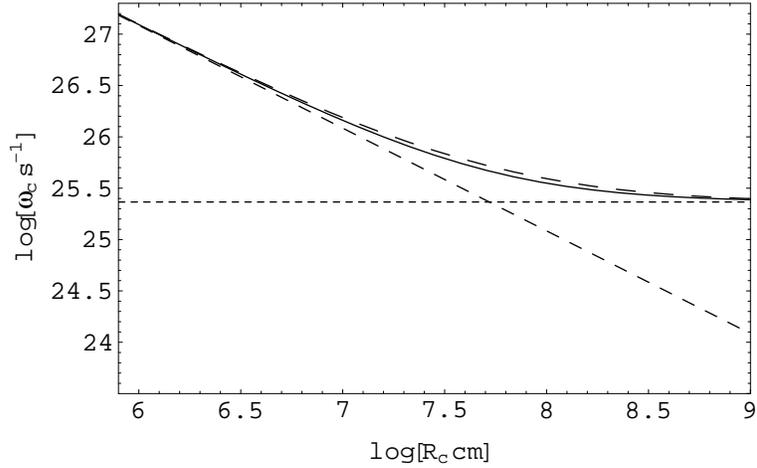}}
\caption{ Comparison (in logarithmic scales) of the variation with the
radius of curvature $R_c$ of the critical effective frequency in the
generalized synchrocurvature model (solid curve) with the critical frequency
of the synchrotron radiation (dotted curve), of the curvature radiation
(short dashed curve) and of the synchrocurvature mechanism proposed by Cheng
\& Zhang (1996) (long dashed curve) for $\sin \alpha =10^{-3}$, $\gamma =3\times 10^7$ and $B=10^6 G$. }
\label{FIG8}
\end{figure}

The more physical approach adopted in the present paper leads to results
very similar to that based on the detailed and complicated mathematical
formalism of the simple synchrocurvature mechanism.

The spectral distribution of the total (over all directions) radiation
intensity $\frac{dP}{d\omega }$ as a function of the frequency $\omega $ is
represented for an ultrarelativistic electron moving in a curved
inhomogeneous magnetic field, for the synchrotron radiation, for the
curvature radiation and for the synchrocurvature mechanism of Cheng \& Zhang
(1996) in Fig. 9.

\begin{figure}[h]
\epsfxsize=10cm \centerline{\epsffile{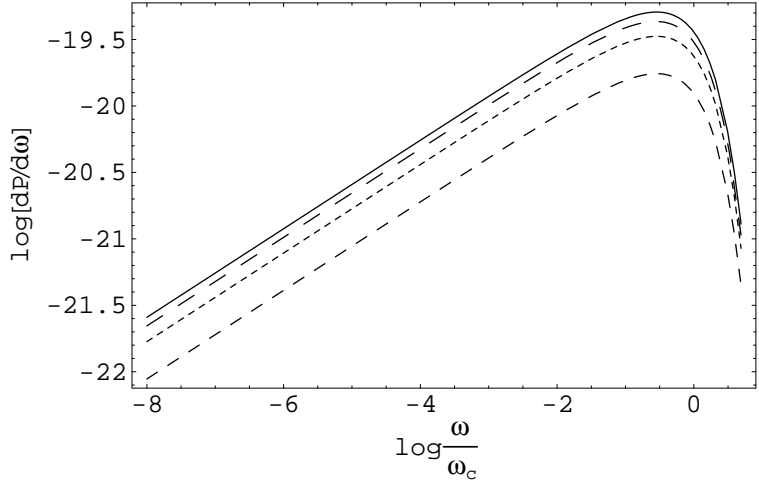}}
\caption{ Spectral distribution of the logarithm of the radiation intensity $%
\log (dP/d\omega )$ as a function of $\log(\omega /\omega
_c)$ emitted by an electron moving in a curved inhomogeneous magnetic field
(solid curve), of the synchrotron radiation (dotted curve), of the curvature
radiation (short dashed curve) and of the synchrocurvature mechanism Cheng
\& Zhang (1996) (long gashed curve) for $\sin \alpha =10^{-3}$, $\gamma =3\times 10^7$, $R_c=10^8 cm$ and $B=10^6 G$. }
\label{FIG9}
\end{figure}

To obtain the total radiation intensity $P(\omega )$ we have to integrate
the spectral distribution of the radiation over $\omega $. The distributions
of the intensity for the generalized synchrocurvature model, the
synchrocurvature model of Cheng \& Zhang (1996), synchrotron radiation and
curvature radiation are presented in Fig. 10.

\begin{figure}[h]
\epsfxsize=10cm \centerline{\epsffile{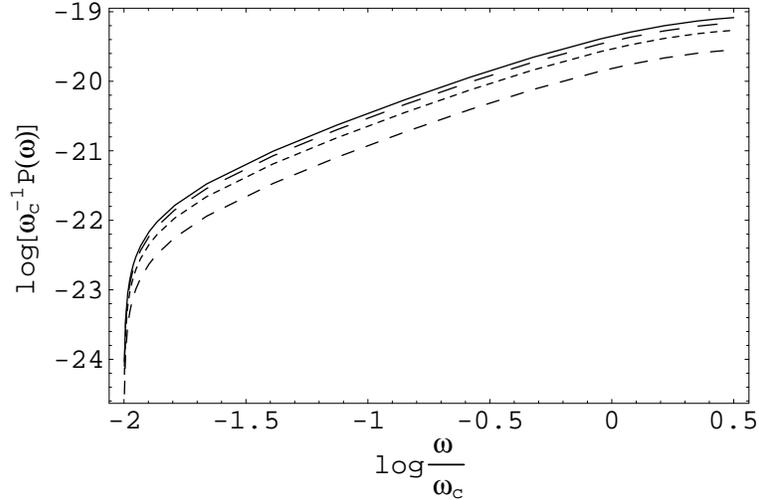}}
\caption{ Variation as a function of $\frac{\omega }{\omega %
_c}$ of the total radiation intensity $P(\omega )$ emitted by an
electron moving in a curved inhomogeneous magnetic field (solid curve), of
the synchrotron radiation (dotted curve), of the curvature radiation (short
dashed curve) and of the synchrocurvature mechanism Cheng \& Zhang (1996)
(long dashed curve) for $\sin \alpha =10^{-3}$, $\gamma %
=3\times 10^7$, $R_c=10^8 cm$ and $B=10^6 G$. }
\label{FIG10}
\end{figure}

In this case also there is an excellent agreement between the results
obtained from Eq. (\ref{p}) and the synchrocurvature model.

The degree of polarization of the total power per unit frequency interval at
a given frequency is defined as 
\begin{equation}
\pi \left( \omega \right) =\frac{K_{2/3}\left( \frac{\omega }{\omega _{c}}%
\right) }{\int_{\omega /\omega _{c}}^{\infty }K_{5/3}\left( \xi \right) d\xi 
}.
\end{equation}

The variation of $\pi \left( \omega \right) $ as a function of frequency for
the generalized synchrocurvature mechanism, synchrotron radiation and
curvature radiation is represented in Fig. 11.

\begin{figure}[h]
\epsfxsize=10cm \centerline{\epsffile{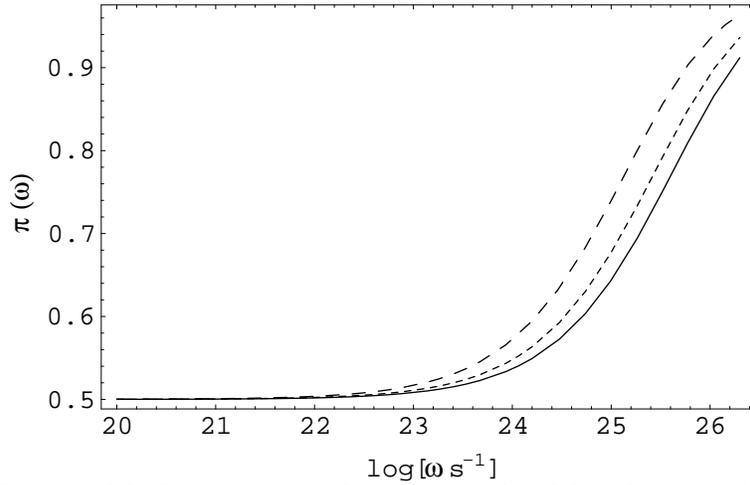}}
\caption{ Variation as a function of the frequency $\omega $ (in a
logarithmic scale) of the polarization degree $\pi (\omega ) 
$ for the generalized synchrocurvature mechanism (solid curve), for the
synchrotron radiation (dotted curve) and for the curvature radiation (dashed
curve) for $\sin \alpha =10^{-3}$, $\gamma =3\times 10^7$, $%
R_c=10^8 cm$ and $B=10^6 G$. }
\label{FIG11}
\end{figure}

As already mentioned and extensively discussed in Cheng \& Zhang (1996), the
degree of polarization can, in principle, provide an effective observational
tool in order to distinguish between the different radiation mechanisms.

The spectral distribution of the radiation for an electron moving in an
inhomogeneous curved magnetic field can also be obtained by using a
phenomenological approach, thus avoiding lengthy derivations of the basic
formulae. Also we obtain  a physical interpretation of the origin of the
radiation spectrum in more complicated magnetic fields. To do this we use the
result that the motion along the curved magnetic field lines and the
gradient of the field can be considered, from a physical point of view, as
an evolution under the action of combined magnetic and electric forces. The
radiation in a given direction occurs mainly from the portion of the
trajectory of the electron in which the velocity of the particle is almost
parallel to that direction. To take into account the effect of the electric
field we define a total electromagnetic force $eF_{eff}$, so that $\left(
eF_{eff}\right) ^{2}$ is the sum of the squares of the transverse components
of the Lorentz force $\vec{F}_{eff}=e\vec{E}_{eff}+\frac{e}{c}\vec{v}%
_{0}\times \vec{B}$ (Landau \& Lifshitz 1975), which can be considered
constant within this segment. Since a small segment of a curve can be
considered as an arc of circle, we can apply the general results obtained
for radiation during uniform motion in a circle in the presence of a
constant magnetic field, replacing $B$ by $F_{eff}$, the value of the
transverse force at a given point.

In particular it follows that the main part of the radiation is concentrated
in the frequency range (Landau \& Lifshitz 1975) 
\begin{equation}  \label{12}
\omega \sim \omega _{Feff}\left( \frac{E}{mc^{2}}\right) ^{3}=\frac{eF_{eff}%
}{mc}\left( \frac{E}{mc^{2}}\right) ^{2}.
\end{equation}

By assuming an ultra-relativistic motion with $v\approx c$ of the charged
particle in the curved inhomogeneous magnetic field, we obtain for $F_{eff}$ 
\begin{equation}  \label{15}
F_{eff}=\sqrt{\vec{B}^{2}+2\vec{B}\cdot \vec{E}_{eff}+\vec{E}_{eff}^{2}}=%
\sqrt{B^{2}\sin ^{2}\alpha +2\frac{\gamma mB}{e}\frac{v_{\mid \mid }^{2}+%
\frac{1}{2}v_{\perp }^{2}}{R_{c}}\sin \alpha +\frac{\gamma ^{2}m^{2}}{e^{2}}%
\frac{\left( v_{\mid \mid }^{2}+\frac{1}{2}v_{\perp }^{2}\right) ^{2}}{%
R_{c}^{2}}}.
\end{equation}

Therefore to obtain the electromagnetic radiation due to the motion in the
combined magnetic and electric field we must substitute the magnetic field $%
B\sin \alpha $ with the effective electromagnetic field in the expressions
of the power and characteristic frequency, corresponding to the radiation in
the presence of a constant magnetic field.

As a first example of this procedure we consider the total power emitted by the
electron in the generalized synchrocurvature model. By substituting the magnetic field
$B\sin \alpha $ in the formula giving the power emitted by synchrotron radiation, Eq. (\ref{synch}),
we immediately obtain Eq. (\ref{a}), the power emitted via the generalized mechanism.

By performing the same substitution in the
expressions of the radiation spectrum and of the characteristic frequency
of the synchrotron radiation, we
obtain the exact radiation spectrum and characteristic frequency of the unified radiation
mechanism, given by Eqs. (\ref{p}) and (\ref{17}). Therefore this physical approach is
consistent with the exact results, obtained without any particular
assumptions.

\section{Generalized synchrocurvature radiation in quantum domain}

In the previous Section we have considered the classical radiation emitted by a
charged particle moving in a curved inhomogeneous magnetic field.
Due to the complicated character of this problem, we have adopted a simple
formalism based on the definition of an effective electromagnetic field $%
F_{eff}$, combining the effects of the space varying magnetic field and of
the electric field associated to the motion of the particle along the
magnetic field lines. The radiation spectrum of the electron, involving the
synchrotron, curvature and emission due to magnetic inhomogeneities
mechanisms, has been obtained by substituting the magnetic field $B$ in the
classical formulae of the synchrotron radiation by $F_{eff}$. The numerical
results obtained by this simple method are in a quite good agreement with
the results obtained by using the formalism of (Cheng \& Zhang 1995; Cheng \& Zhang 1996), based
on a complicated mathematical approach.

It is the purpose of the present Section to extend this method to the study
of the quantum effects arising in the radiation of an electron moving in a
curved inhomogeneous magnetic field.

In the quantum theory radiation intensity differs from that in the classical
theory and is transformed to it as $\hbar \rightarrow 0$. There are two
types of quantum effects taking place, namely an effect due to the quantum
character of particle motion in an external field and that due to the recoil
that a particle undergoes during radiation. These effects are interrelated
but at high energies they can be separated. The former effect, which can be
described as one due to the noncomutativity of operators compared to
classical values (i.e. operators of the particle momentum, energy, velocity
etc.), appears to be less essential than the recoil effect during radiation.

There are several methods of obtaining the quantum corrections to the
radiation spectrum of the synchrotron radiation like the Fermi-Weizsacker-Williams (FWW) virtual photons
method (for early references and a discussion of recent results see Zolotorev \& McDonald (2000)), Schwinger's operator method (Schwinger, Tsai \& Erber 1976; Akhiezer \& Shulga 1996)
and the semiclassical approximation (Berestetskii, Lifshitz \& Pitaievskii 1989).
The FWW approach is based on the observation that the electromagnetic fields
of an electron in uniform relativistic motion are predominantly transverse, with $\vec{E}\approx \vec{B}$.
This is very much like the fields of a plane wave, so one is led to regard the
fast electron as carrying with it a cloud of virtual photons that it can radiate
if perturbed. In the operator method the first quantum correction is involved by the
substitution $\omega \rightarrow \omega \left( 1+\frac{\hbar \omega }{E}%
\right) $ ( $E$ is the particle energy) in the classical expression for $%
\omega ^{-1}\frac{dI}{d\omega }$ (Akhiezer \& Shulga 1996). Therefore in this formalism the spectral density can be
written as
\begin{equation}\label{cor}
\frac{dI}{d\omega }=\frac{1}{\sqrt{3}\pi }\frac{e^{2}}{c\gamma ^{2}}\omega
\int_{\left( 1+\frac{\hbar \omega }{E}\right) \omega }^{\infty
}K_{5/3}\left( \xi \right) d\xi. 
\end{equation}

The quantum correction to the total loss of particle energy in a magnetic
field is defined by $\frac{\hbar \gamma ^{2}}{mR}, R=\frac{E}{eB}$. This
correction increases with the particle energy and the magnetic field
strength and at high energies and strengths can become significant.

In the following discussion we shall not adopt the operator method but
instead we use the quasi-classical method as described in detail in Berestetskii,
Lifschitz \& Pitayevskii (1989). In this approach to the problem of the quantum radiation
in strong magnetic fields it is explicitly used the
fact that the motion of the particle is quasi-classical. The quantum recoil
associated with the emission of a photon is determined by the ratio $\frac{%
\hbar \omega }{E}$. For the classical theory to be applicable this parameter
must be small. In order to take into account quantum effects it is
convenient to use the parameter $\chi =\frac{B}{B_{0}}\frac{\left| \vec{p}%
\right| }{mc}=\frac{B}{B_{0}}\frac{E}{mc^{2}}=\frac{\hbar \omega _{B}}{E}%
\left( \frac{E}{mc^{2}}\right) ^{3}$, where $B_{0}=\frac{m^{2}c^{3}}{e\hbar 
}=4.4\times 10^{13}G$ and $\omega _{B}=\frac{eBv}{pc}=\frac{eBc}{E}$. In
the classical case $\chi \sim \frac{\hbar \omega }{E}<<1$. When
quantum effects become important, the energy of the emitted photon $\hbar
\omega \sim E$ and $\chi >>1$. In this case the significant region
of the spectrum extends to frequencies at which the electron energy after
emission is $E^{\prime }\approx mc^{2}\frac{B_{0}}{B}$.

We assume that the electron is ultra-relativistic and the
magnetic field satisfies the condition $\frac{B}{B_{0}}<1$. This condition
is satisfied with a very good approximation for magnetic fields of the order 
$10^{12}G$. The interval between adjacent energy levels for motion in a
magnetic field is $\hbar \omega _{B}$ and the quantization of the electron
energy is expressed by the ratio $\frac{\hbar \omega _{B}}{E}=\left( \frac{B%
}{B_{0}}\right) \left( \frac{mc^{2}}{E}\right) ^{2}$. It follows then that $%
\hbar \omega _{B}<<E$, that is the motion of the electron is
quasi-classical for all values of $\chi $.

The probability of emission of the photon can be calculated by using
perturbation theory (the detailed calculations are presented in
Berestetskii, Lifschitz \& Pitayevskii (1989). With the inclusion of the
quantum effects the spectral distribution of the radiation is given by
\begin{equation}\label{19}
\frac{dI}{d\omega }=-\frac{e^{2}m^{2}c^{4}\omega }{\sqrt{\pi }E^{2}}\left[
\int_{x}^{\infty }\Phi \left( \xi \right) d\xi +\left( \frac{2}{x}+\frac{%
\hbar \omega }{E}\chi \sqrt{x}\right) \Phi ^{\prime }\left( x\right) \right]
,
\end{equation}
where $\Phi $ is the Airy function and $x=\left( \frac{\hbar \omega }{%
E^{\prime }\chi }\right) ^{\frac{2}{3}}=\frac{m^{2}c^{4}}{E^{2}}\left( \frac{%
E\omega }{E^{\prime }\omega _{B}}\right) ^{\frac{2}{3}}$. In the classical
limit $\hbar \omega <<E^{\prime }\approx E$ and the variable $x$ becomes $%
x=\left( \frac{\omega }{\omega _{B}}\right) ^{\frac{2}{3}}\left( \frac{mc^{2}%
}{E}\right) ^{2}$. Hence the second term in the round brackets is small and we
obtain the classical formula (Landau \& Lifshitz 1975).

The critical frequency is defined in the quantum case according to $\hbar
\omega _{c}=\frac{E\chi }{\frac{2}{3}+\chi }$ (Berestetskii, Lifschitz \& Pitayevskii 1989).
In the limit $\chi <<1$ we obtain again the classical critical frequency for the synchrotron radiation $%
\omega _{c}=\frac{3eB\sin \alpha }{2mc}\left( \frac{E}{mc^{2}}\right) ^{2}$ (Landau and Lifshitz 1975).

To calculate the total radiation intensity we must integrate Eq. (\ref{19})
with respect to $\omega $ from $0$ to $E$. The final result is 
(Berestetskii, Lifschitz \& Pitayevskii 1989)
\begin{equation}\label{20}
I=-\frac{e^{2}m^{2}c^{4}\chi ^{2}}{2\sqrt{\pi }\hbar ^{2}}\int_{0}^{\infty }%
\frac{4+5\chi x^{\frac{3}{2}}+4\chi ^{2}x^{3}}{\left( 1+\chi x^{\frac{3}{2}%
}\right) ^{4}}\Phi ^{\prime }(x)xdx.
\end{equation}

There are two important limiting cases of Eq. (\ref{20}). In the classical
limit $\chi <<1$ we obtain
$I=\frac{2e^{4}B^{2}\sin ^{2}\alpha E^{2}}{3m^{4}c^{7}}\left( 1-\frac{55%
\sqrt{3}}{16}\chi +48\chi ^{2}-...\right) $. In the opposite limit the
result is $I=0.37\frac{e^{2}m^{2}}{\hbar ^{2}}\left( \frac{B\sin \alpha }{%
B_{0}}\frac{E}{mc^{2}}\right) ^{\frac{2}{3}}$.

In order to find the general form of the dependence of the radiation
intensity emission for an electron in a constant electromagnetic field, we
note first that the state of a particle in any constant and uniform field is
defined by as many quantum numbers as the state of a free particle, and this
may always be so chosen as to become, when the field is removed, those of a
free particle, i.e. its 4-momentum $p^{\mu }$ ($p^{2}=m^{2}c^{2}$ ). Thus
the state of a particle is described in a constant electromagnetic field by
a constant 4-vector $p$. The total intensity of emission, being an
invariant, depends only on the invariants which can be constructed from the
constant 4-tensor $F_{\mu \nu }$ and the constant 4-vector $p^{\mu }$.

Since $F_{\mu \nu }$ can appear in the intensity only in combination with
the charge $e$, there are three dimensionless invariants that can be
formed: $f_{1}^{2}=-\frac{2}{3}\frac{e^{4}}{m^{4}c^{5}}\left( F_{\mu \nu
}p^{\nu }\right) ^{2}$,  $f_{2}^2=\frac{e^{4}}{m^{4}c^{5}}\left( F_{\mu \nu
}\right) ^{2}$ and $f_{3}^2=\frac{e^{4}}{m^{4}c^{5}}e_{\lambda \mu \nu \rho
}F^{\lambda \mu }F^{\nu \rho }$, with $e_{\lambda \mu \nu \rho }$ the
totally antisymmetric tensor. But it can be shown (Berestetskii, Lifshitz
\& Pitaevski 1989) that if the electron is ultra-relativistic then
$f_{1}^{2}>>f_{2}^2, f_{3}^2$. If also the fields $\left| \vec{E}\right| $, $\left| \vec{B}%
\right| <<m^{2}c^{3}/e\hbar =B_{0}$, then $\left| f_{2}^2\right| ,\left| f_{3}^2\right|
<<1$ and the intensity of the emission in any constant field is expressed
only in terms of the invariant $f_{1}^{2}=\frac{2}{3}\frac{e^{4}}{m^{4}c^{5}}%
\left[ \left( \vec{p}\times \vec{B}+p_{0}\vec{E}\right) ^{2}-\left( \vec{p}%
\cdot \vec{E}\right) ^{2}\right] $. With $p_{0}=\gamma mc$, $\vec{p}%
=\gamma m\vec{v}_{0}$ and assuming an ultrarelativistic motion of the electron $%
\vec{v}_{0}\approx c\vec{n}$ we obtain 
$f_{1}^{2}=\frac{2}{3}\frac{e^{4}}{m^{2}c^{3}}\gamma \left[ ^{2}\left( \vec{n%
}\times \vec{B}+\vec{E}\right) ^{2}-\left( \vec{p}\cdot \vec{E}\right) ^{2}%
\right] $.

For $\vec{E}=0$ we find $f_{1}^{2}=\frac{2}{3}\frac{e^{4}B^{2}\sin ^{2}\alpha }{m^{2}c^{3}}\gamma ^{2}$, that
is in this case the invariant $f_{1}^{2}$ gives the total intensity of the synchrotron
radiation in the ultrarelativistic case.

We generalize now the previous results to the case of the radiation emitted
by an electron moving in a curved inhomogeneous magnetic field. As we have seen, the total intensity
of the radiation must be a function of the invariant $f_{1}^{2}$ only. Therefore   
we can use the method applied in the previous Section for the study of the classical
process, that is, we substitute in all formulae above the magnetic field
with the effective electromagnetic field $F_{eff}$, including the
fictitious electric field which can replace the effect of the centrifugal
force determining the motion of the electron along the magnetic field lines
and of the transverse motion due to the gradient of the field. In fact the generalized
electromagnetic field $F_{eff}$ can also be obtained from the expression of the
invariant $f_{1}^2$ by taking into account that the velocity $\vec{v}_{0}$ of the electron,
corresponding to the motion in the constant magnetic field and the radius of curvature $\vec{R}_{c}$
are perpendicular, $\vec{v}_{0}\cdot \vec{R}_{c}=0$ and hence $\vec{p}\cdot \vec{E}=0$.

Hence for the critical frequency of the generalized synchrocurvature 
radiation emitted in the quantum domain we obtain
\begin{equation}\label{crit}
\omega _{c}=\frac{\frac{e}{mc}\left( \frac{E}{mc^{2}}\right) ^{2}\sqrt{%
B^{2}\sin ^{2}\alpha +2\frac{\gamma mB}{e}\frac{v_{\mid \mid }^{2}+\frac{1}{2%
}v_{\perp }^{2}}{R_{c}}\sin \alpha +\frac{\gamma ^{2}m^{2}}{e^{2}}\frac{%
\left( v_{\mid \mid }^{2}+\frac{1}{2}v_{\perp }^{2}\right) ^{2}}{R_{c}^{2}}}%
}{\frac{2}{3}+\frac{\hbar e}{m^{2}c^{3}}\frac{E}{mc^{2}}\sqrt{B^{2}\sin
^{2}\alpha +2\frac{\gamma mB}{e}\frac{v_{\mid \mid }^{2}+\frac{1}{2}v_{\perp
}^{2}}{R_{c}}\sin \alpha +\frac{\gamma ^{2}m^{2}}{e^{2}}\frac{\left( v_{\mid
\mid }^{2}+\frac{1}{2}v_{\perp }^{2}\right) ^{2}}{R_{c}^{2}}}}.  
\end{equation}

In the limit $\hbar \to 0$ we obtain again Eq. (\ref{17}), the characteristic frequency of the radiation
in the clasical domain.
The variation of the quantum critical frequency as a function of $\sin \alpha $, of the
particle energy, magnetic field and radius of curvature is represented in Figs. 12-15.
For the sake of comparison we have also presented the variation of the classical critical
frequency of the generalized synchrocurvature radiation mechanism in the same domain of 
physical parameters.

\begin{figure}[h]
\epsfxsize=10cm
\centerline{\epsffile{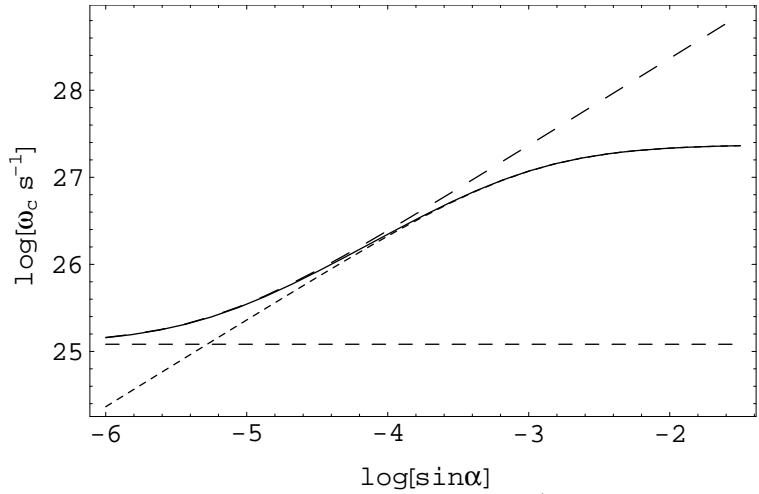}}
\caption{
Variation in the quantum domain of the critical frequency $\omega _c (s^{-1})$ , as a function of $\sin \alpha $ (in logarithmic scales), 
for the generalized synchrocurvature mechanism
(solid curve), for the synchrotron radiation (Berestetskii et al. 1989) (dotted curve) and for the curvature radiation (dashed curve),
as compared with the classical critical frequency of the generalized synchrocurvature mechanism (long dashed curve),
for $\gamma =3\times 10^6$, $R_c=10^5 cm$ and $B=10^{10} G$. 
}
\label{FIG12}
\end{figure}

\begin{figure}[h]
\epsfxsize=10cm
\centerline{\epsffile{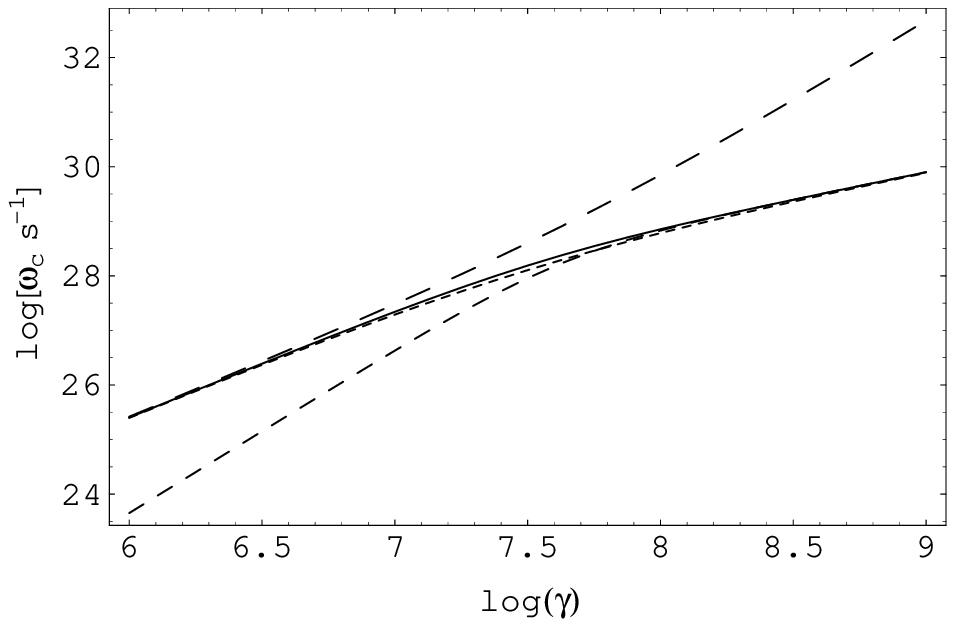}}
\caption{
Variation in the quantum domain of the critical frequency $\omega _c (s^{-1})$, as a function of $\gamma $ (in logarithmic scales), 
for the generalized synchrocurvature mechanism
(solid curve), for the synchrotron radiation (Berestetskii et al. 1989) (dotted curve) and for the curvature radiation (dashed curve), as
compared with the classical critical frequency of the generalized synchrocurvature mechanism (long dashed curve),
for $\sin \alpha =10^{-3}$, $R_c=10^5 cm$ and $B=10^9 G$. 
}
\label{FIG13}
\end{figure}

\begin{figure}[h]
\epsfxsize=10cm
\centerline{\epsffile{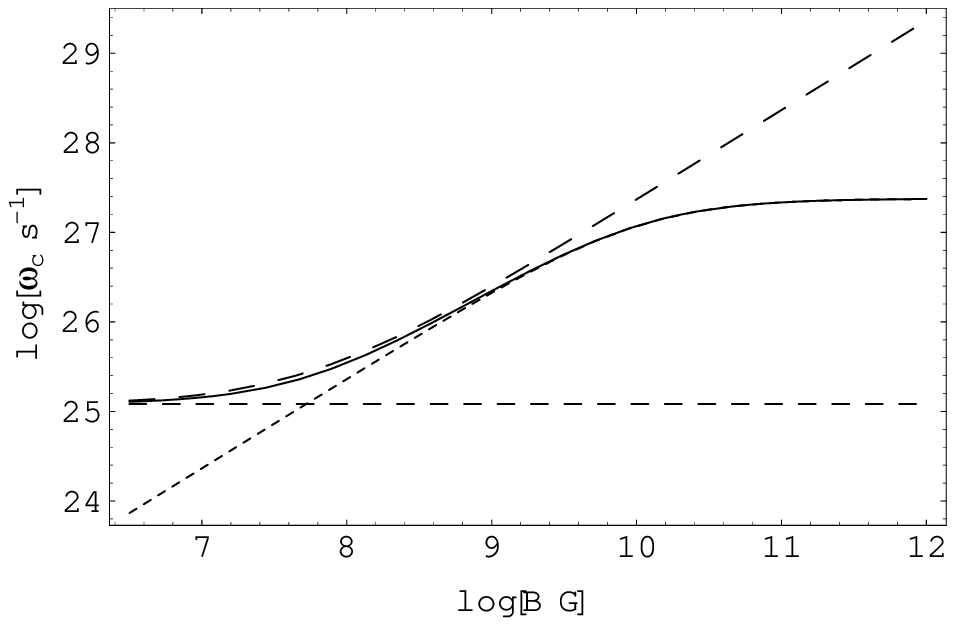}}
\caption{
Variation in the quantum domain of the critical frequency $\omega _c (s^{-1})$, as a function of the magnetic field $B(G)$ (in logarithmic scales), 
for the generalized synchrocurvature mechanism
(solid curve), for the synchrotron radiation (Berestetskii et al. 1989) (dotted curve) and for the curvature radiation (dashed curve),
as compared with the classical critical frequency of the generalized synchrocurvature mechanism (long dashed curve),
for $\sin \alpha =10^{-3}$, $R_c=10^5 cm$ and $\gamma =3\times 10^6$.
}
\label{FIG14}
\end{figure}

\begin{figure}[h]
\epsfxsize=10cm
\centerline{\epsffile{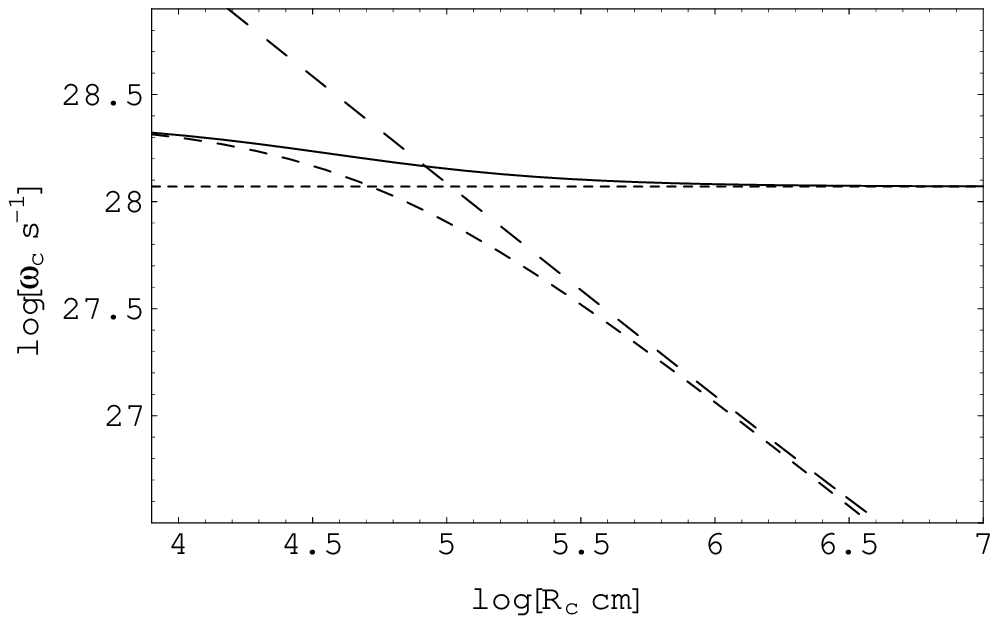}}
\caption{
Variation in the quantum domain of the critical frequency $\omega _c (s^{-1})$, as a function of the radius of curvature $R_{c} (cm)$ (in logarithmic scales), 
for the generalized synchrocurvature mechanism
(solid curve), for the synchrotron radiation (Berestetskii et al. 1989) (dotted curve) and for the curvature radiation (dashed curve),
as compared with the classical critical frequency of the generalized synchrocurvature mechanism (long dashed curve),
for $\sin \alpha =10^{-3}$, $B=10^9 G$ and $\gamma =3\times 10^7$.
}
\label{FIG15}
\end{figure}

The frequency distribution for the generalized synchrocurvature radiation
mechanism is given by
\begin{eqnarray}
\frac{dI}{d\omega } &=&-\frac{e^{2}m^{2}c^{4}\omega }{\sqrt{\pi }E^{2}}%
\times  \nonumber \\
&&\left[ \int_{x}^{\infty }\Phi \left( \xi \right) d\xi +\left( \frac{2}{x}+%
\frac{\hbar \omega }{B_{0}}\frac{\sqrt{B^{2}\sin ^{2}\alpha +2%
\frac{\gamma mB}{e}\frac{v_{\mid \mid }^{2}+\frac{1}{2}v_{\perp }^{2}}{R_{c}}%
\sin \alpha +\frac{\gamma ^{2}m^{2}}{e^{2}}\frac{\left( v_{\mid \mid }^{2}+%
\frac{1}{2}v_{\perp }^{2}\right) ^{2}}{R_{c}^{2}}}}{mc^{2}}\sqrt{x}\right)
\Phi ^{\prime }\left( x\right) \right],\nonumber\\
\end{eqnarray}
where the variable $x$ is defined as $x=\left( \frac{m^{2}c^{4}}{E^{2}}\right)
\left( \frac{\omega E^{2}}{emc^{3}B_{0}}\right) $. The variations of $\frac{dI}{d\omega }$
for the generalized synchrocurvature radiation, for the synchrotron radiation and for
the curvature radiation are presented, for different values of the pitch angle $\alpha $, in Figs. 16 and 17.

\begin{figure}[h]
\epsfxsize=10cm
\centerline{\epsffile{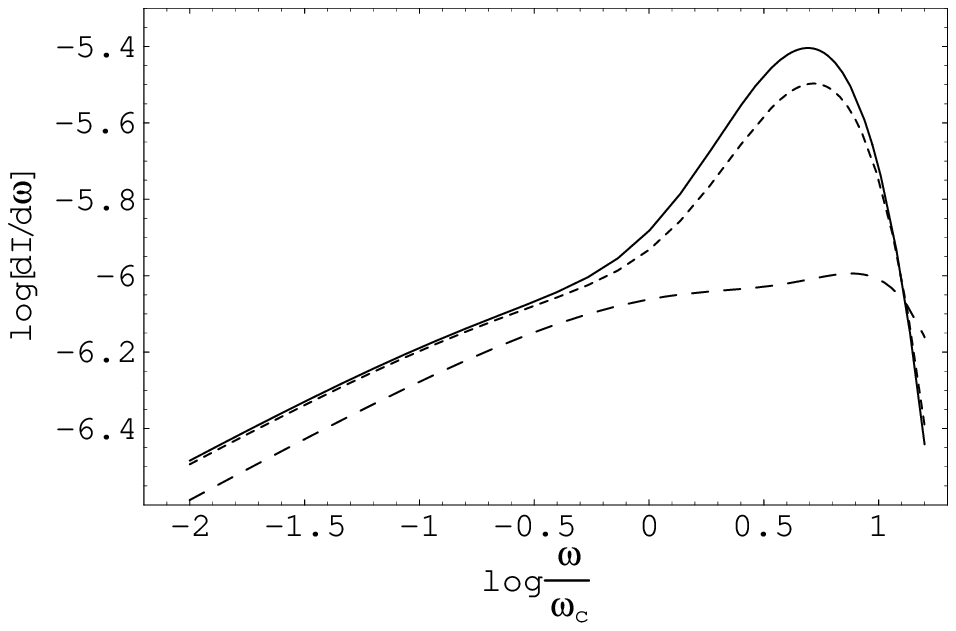}}
\caption{
Variation, in the quantum domain, of the frequency distribution $\log(dI/d\omega )$ as a function of the frequency $\log(\omega /\omega _c)$ (in logarithmic scales),
for the generalized synchrocurvature mechanism (solid curve), for the synchrotron radiation (Berestetskii et al. 1989) (dotted curve) and for
the curvature radiation (dashed curve),
for $\sin \alpha =2\times 10^{-3}$, $B=10^9 G$, $\gamma =3\times 10^7$ and $R_c=10^5cm$.
}
\label{FIG16}
\end{figure}

\begin{figure}[h]
\epsfxsize=10cm
\centerline{\epsffile{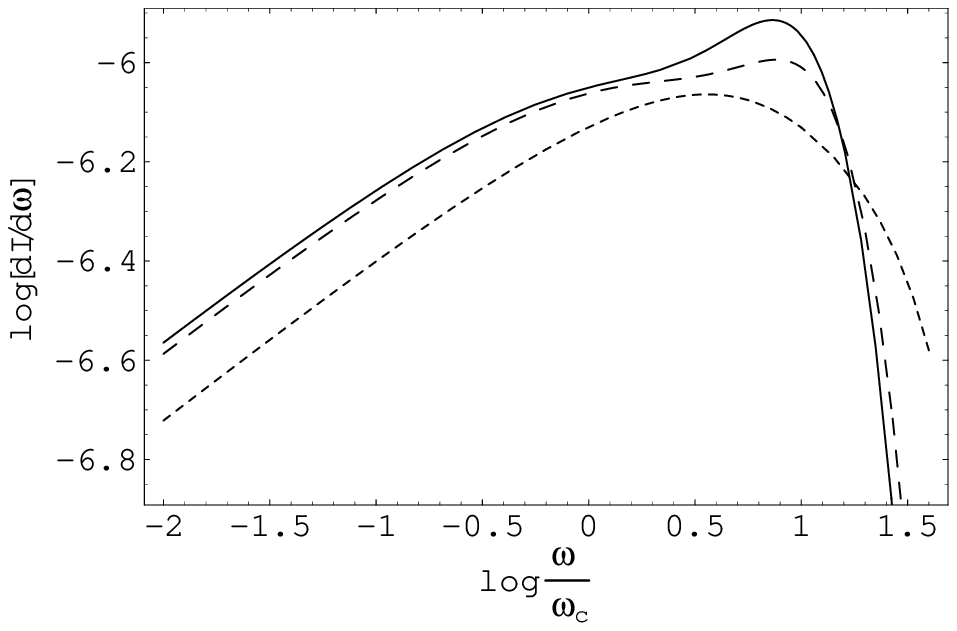}}
\caption{
Variation, in the quantum domain, of the frequency distribution $\log(dI/d\omega )$ as a function of the frequency $\log(\omega /\omega _c)$ (in logarithmic scales),
for the generalized synchrocurvature mechanism (solid curve), for the synchrotron radiation (Berestetskii et al. 1989) (dotted curve) and for
the curvature radiation (dashed curve),
for $\sin \alpha =2\times 10^{-5}$, $B=10^9 G$, $\gamma =3\times 10^7$ and $R_c=10^5cm$.
}
\label{FIG17}
\end{figure}

For the total radiation intensity in the quantum domain we find
\begin{equation}\label{crit2}
I=-\frac{e^{2}m^{2}c^{4}\chi _{eff}^{2}}{2\sqrt{\pi }\hbar ^{2}}%
\int_{0}^{\infty }\frac{4+5\chi _{eff}x^{\frac{3}{2}}+4\chi _{eff}^{2}x^{3}}{%
\left( 1+\chi _{eff}x^{\frac{3}{2}}\right) ^{4}}\Phi ^{\prime }(x)xdx,
\end{equation}
where we denoted
\begin{equation}
\chi _{eff}=\frac{E}{mc^{2}}\frac{\sqrt{B^{2}\sin ^{2}\alpha +2%
\frac{\gamma mB}{e}\frac{v_{\mid \mid }^{2}+\frac{1}{2}v_{\perp }^{2}}{R_{c}}%
\sin \alpha +\frac{\gamma ^{2}m^{2}}{e^{2}}\frac{\left( v_{\mid \mid }^{2}+%
\frac{1}{2}v_{\perp }^{2}\right) ^{2}}{R_{c}^{2}}}}{B_{0}}.
\end{equation}

The approximate values of the total
intensity of the generalized synchrocurvature mechanism in the two limits, the classical and the quantum, respectively,
can also be obtained with the help of the substitution $\chi \rightarrow
\chi _{eff}$. The total radiation intensity $I$ versus the magnetic field $B$,
$\gamma $, $\sin \alpha$ and the radius of curvature $R_c$ is presented in Figs. 18-21.

\begin{figure}[h]
\epsfxsize=10cm
\centerline{\epsffile{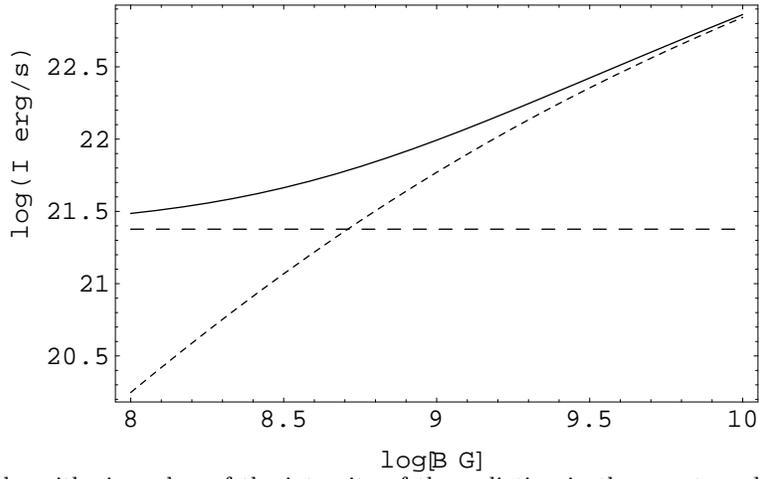}}
\caption{
Variation, in logarithmic scales, of the intensity of the radiation in the quantum domain, as a function of the magnetic
field $B(G)$,  for the generalized synchrocurvature mechanism (solid curve), for the synchrotron radiation (Berestetskii et al. 1989) (dashed curve),
for $\sin \alpha =10^{-3}$, $\gamma =3\times 10^7$ and $R_c=10^5 cm$. 
}
\label{FIG18}
\end{figure}

\begin{figure}[h]
\epsfxsize=10cm
\centerline{\epsffile{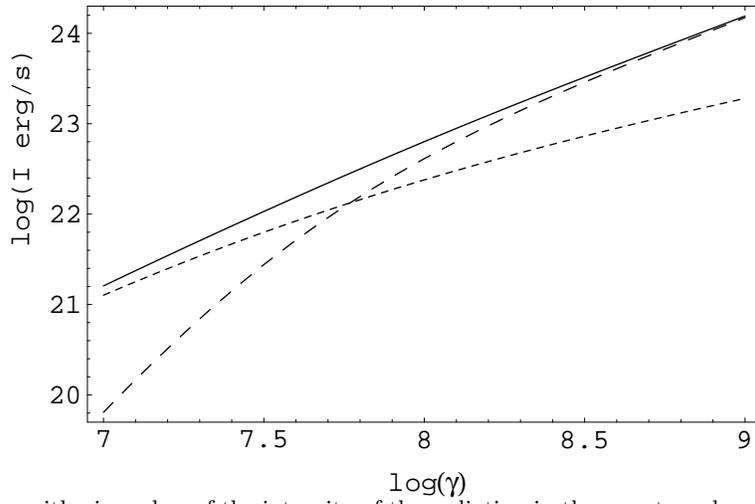}}
\caption{
Variation, in logarithmic scales, of the intensity of the radiation in the quantum domain as a function of $\gamma $,
for the generalized synchrocurvature mechanism (solid curve),
for the synchrotron radiation (Berestetskii et al. 1989) (dotted curve) and for the curvature radiation (dashed curve)
for $\sin \alpha =10^{-3}$, $B=10^9 G$ and $R_c=10^5 cm$. 
}
\label{FIG19}
\end{figure}

\begin{figure}[h]
\epsfxsize=10cm
\centerline{\epsffile{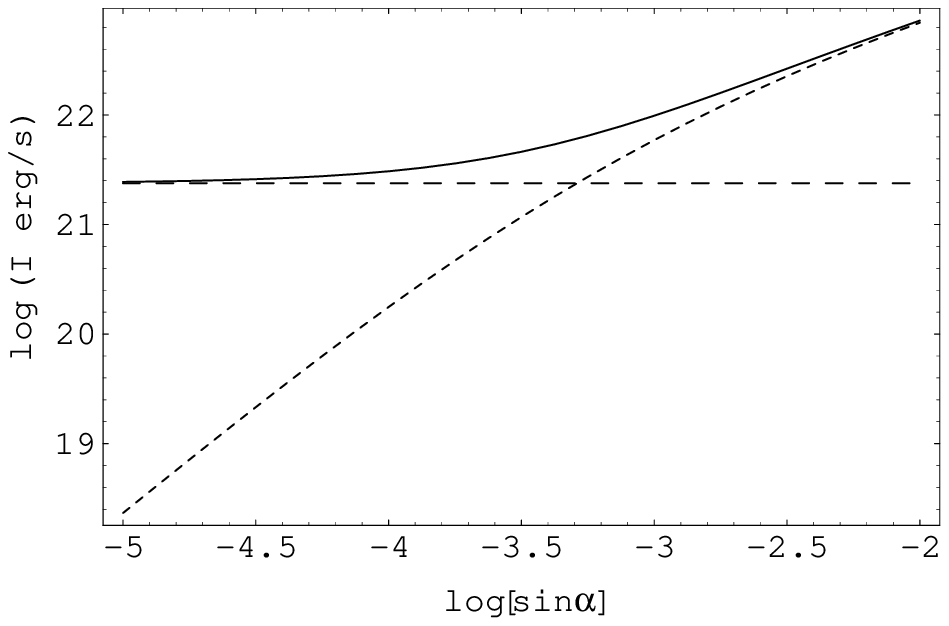}}
\caption{
Variation, in logarithmic scales, of the intensity of the radiation in the quantum domain as a function of 
$\sin \alpha $, for the generalized synchrocurvature mechanism (solid curve),
for the synchrotron radiation (Berestetskii et al. 1989) (dotted curve) and for the curvature radiation (dashed curve)
for $\gamma =3\times 10^6$, $B=10^9 G$ and $R_c=10^5 cm$. 
}
\label{FIG20}
\end{figure}

\begin{figure}[h]
\epsfxsize=10cm
\centerline{\epsffile{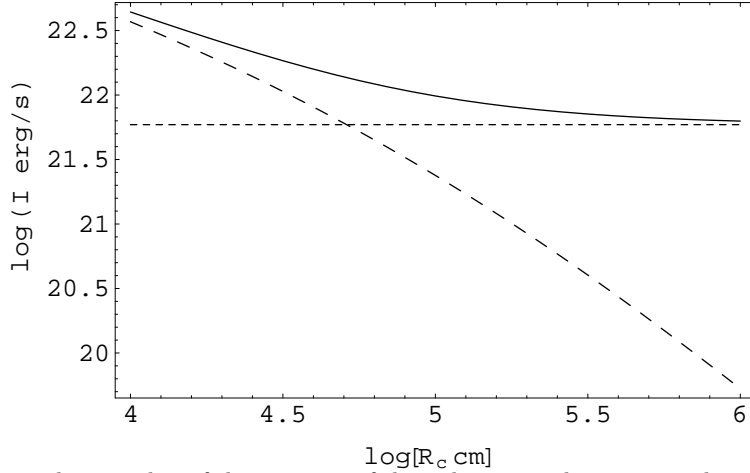}}
\caption{
Variation, in logarithmic scales, of the intensity of the radiation in the quantum domain as a function of 
the radius of curvature $R_c (cm)$, for the generalized synchrocurvature mechanism (solid curve),
for the synchrotron radiation (Berestetskii et al. 1989) (dotted curve) and for the curvature radiation (dashed curve)
for $\gamma =3\times 10^7$, $B=10^9 G$ and $\sin \alpha =10^{-3}$. 
}
\label{FIG21}
\end{figure}

\section{Effects of magnetic field inhomogeneities on radiation transitions rates} 

In a uniform homogeneous magnetic field electrons must occupy discrete
energy states 
\begin{equation}
E_{n}=mc^{2}\left[ 1+\left( \frac{p_{\mid \mid }}{mc}\right) ^{2}+\left( 
\frac{p_{D}}{mc}\right) ^{2}+\left( \frac{p_{\perp }}{mc}\right) ^{2}\right]
^{\frac{1}{2}},
\end{equation}
where $p_{\mid \mid }$, $p_{D}$ and $p_{\perp }$ are the momentum components
parallel to the magnetic field direction, the magnetic drift momentum and
momentum perpendicular to the field, respectively (Boussard 1984; Latal 1986). The
perpendicular component is quantized according to $p_{\perp }^{2}=2m\hbar
\omega _{c}\left[ l+\frac{1}{2}\left( s+1\right) \right] =2n\gamma \frac{%
B\sin \alpha }{B_{0}}$, with $l=0,1,2,...$ and $s=+1$ and $s=-1$ for the
spin-up and spin-down states, respectively. In the ground state $n=0$ only
the spin-down state is allowed. The pitch angle $\alpha $ is generally
quantized according to $\tan \alpha =p_{\perp }/p_{\mid \mid }$. In the
following we denote $p_{S}^{2}=p_{\mid \mid }^{2}+p_{D}^{2}$. Hence the
state of the electron is completely described by three quantities $\left(
n,s,p_{S}\right) $. As in the previous Sections, we assume that the drift is
due to the combined effects of the curvature and gradient of the magnetic
field and therefore the inhomogeneity and curvature of the magnetic field
will cause a drift with velocity $\vec{v}_{D}=\frac{\gamma mc}{e}\frac{%
v_{\mid \mid }^{2}+\frac{1}{2}v_{\perp }^{2}}{R_{c}^{2}B^{2}}\vec{R}%
_{c}\times \vec{B}$. The corresponding momentum is $p_{D}=\gamma mv_{D}=%
\frac{1}{2}\frac{c\gamma ^{2}m^{2}\left( v_{\mid \mid }^{2}+\frac{1}{2}%
v_{\perp }^{2}\right) }{eR_{c}B}$. 

In order to take into account the effects of the inhomogeneitis of the
magnetic field we assume that these effects can be formally described by
means of the formal substitution $B\rightarrow F_{eff}$, where $F_{eff}$ is
the fictitious electromagnetic field combining the effects of the constant
magnetic field and of the electric field associated with the curvature and
gradient effects.

Hence the energy of the electron moving in an inhomogeneous magnetic field
is quantized according to 
\begin{equation}
E_{n}=mc^{2}\left\{ 1+\left( \frac{p_{\mid \mid }}{mc}\right) ^{2}+\left[ 
\frac{1}{2}\frac{\gamma ^{2}m}{eR_{c}B}\left( v_{\mid \mid }^{2}+\frac{1}{2}%
v_{\perp }^{2}\right) \right] ^{2}+2n\frac{F_{eff}}{B_{0}}\right\} ^{\frac{1%
}{2}}.
\end{equation}

Transitions between the states $\left( n,s,p_{S}\right) $ and $\left(
n^{\prime },s^{\prime },p_{S}^{\prime }\right) $, where $n^{\prime }<n$,
result in the emissions of photons at angle $\theta $ to the magnetic field
with energy $\omega $ given from the kinematics by (Latal 1986; Harding \& Preece 1987)
\begin{equation}\label{em1}
\hbar \omega _{nn^{\prime }}=\frac{2m^{2}c^{4}\left| n-n^{\prime }\right|
B/B_{0}}{\left( 1+\xi \right) \left( E_{n}-cp_{S}\cos \theta \right) },
\end{equation}
where $\xi =\left[ 1-2\left( n-n^{\prime }\right) \frac{B}{B_{0}}%
m^{2}c^{4}\sin ^{2}\theta \left( E_{n}-cp_{S}\cos \theta \right) ^{-2}\right] ^{1/2}$
.

Unless the photon is emitted with angle $\theta =90^{\circ }$, the electron
will experience a recoil along the field direction with a final parallel
momentum $p_{S}^{\prime }=p_{S}-\zeta \omega \cos \theta $, where $s=\left(
n-n^{\prime }\right) /\left| n-n^{\prime }\right| $. For emission $%
n>n^{\prime }$  and $\zeta =+1$ while for absorption ($n<n^{\prime }$ ) $%
\zeta =-1$.

The transition probability per unit time $\Gamma $ between arbitrary initial
and final states and the associated energy rate $\frac{dI}{dt}$ involving
the emission or absorption of a photon of momentum $k^{\mu }=\left( k_{0}=%
\frac{\omega }{c},\vec{k}\right) $ can be expressed in terms over the
current sources in the form $\Gamma =\left( \hbar c\right) ^{-1}\int \frac{%
d^{3}k}{2k_{0}\left( 2\pi \right) ^{3}}\left[ J\right] ^{2}$ and $\frac{dI}{%
dt}=\int \frac{d^{3}k}{2\left( 2\pi \right) ^{3}}\left[ J\right] ^{2}$,
where $\left[ J\right] ^{2}$ represents the invariant magnitude of the
four-dimensional Fourier transform of the transition current,
\begin{equation}
\left[J\right] ^{2}=\sum_{q}\left[ \left| J^{\ast }\left( k\right) \cdot J\left(
k\right) \right| -\left| J_{0}(k)\right| ^{2}\right],
\end{equation}
and $J_{\mu}(k)=ie\int d^{4}x\bar{\psi}_{f}\left( x\right) \gamma _{\mu }\bar{\psi}%
_{i}\left( x\right) e^{-ikx}$ (Latal 1986).

The transition probabilities and the energy rate can be calculated for
transitions between states with arbitrary $n$ and $n^{\prime }$ (Daugherty \& Ventura 1978; Latal 1986). Due to their complicated mathematical
character we do not present them here. In the following we restrict our
analysis only to emission of radiation due to transitions to the ground
state, with $n^{\prime }=0$. In this case the transition rate to the ground
state is (Latal 1986)
\begin{equation}\label{em2}
\Gamma _{n,0}^{(s)}=\left[ 1-s\left( 1+2n\gamma \frac{B\sin \alpha }{B_{0}}%
\right) ^{-1/2}\right] \bar{\Gamma}_{n,0},
\end{equation}
where 
\begin{eqnarray}\label{em3}
&\bar{\Gamma}_{n,0}=\tau _{0}^{-1}\frac{n^{n}}{\left( n-1\right) !}\frac{%
B\sin \alpha }{B_{0}}\frac{\gamma mc^{2}}{E_{n}}\times \nonumber\\
&\int_{0}^{\pi }\sin \theta
\left( \frac{mc^{2}}{E_{n}-cp_{S}\cos \theta }\right) ^{2}\frac{\left( 1-\xi \right)
^{n-1}}{\left( 1+\xi \right) ^{n+1}}\left( 2n\gamma \frac{B\sin \alpha }{%
B_{0}}-\frac{1-\xi }{\xi }\right) \exp \left( -n\frac{1-\xi }{1+\xi }\right)
d\theta,\nonumber\\
\end{eqnarray}
with $\xi $ defined as 
\begin{equation}
\xi =\left[ 1-2n\frac{B\sin \alpha }{B_{0}}\frac{m^{2}c^{4}\sin
^{2}\theta }{\left( E_{n}-cp_{S}\cos \theta \right) ^{2}}\right] ^{1/2},
\end{equation}
and $\tau _{0}=\hbar /\left( \alpha mc^{2}\right) \approx 1.765\times
10^{-19}s$ (in this expression $\alpha $ is the fine structure constant).

The associated energy rate is obtained from $\frac{dI_{n,0}^{(s)}}{dt}=\hbar
\omega \Gamma _{n,0}^{(s)}$.

For the case of the electron moving in the curved inhomogeneous field we
assume that the emission of the radiation from a higher Landau state $%
n=1,2,...$ to the ground state $n=0$ can also be described with the use of
equations (\ref{em1}) and (\ref{em2})-(\ref{em3}) after performing the substitution $B\rightarrow F_{eff}$. 

In Figs. 22-28 we present the variation of $\frac{dI}{dt}$ for an electron
moving in a curved inhomogeneous magnetic field
with respect to the magnetic field $B$, $\gamma $ and the radius of curvature $R_c$
for different values of $n$.

\begin{figure}[h]
\epsfxsize=10cm
\centerline{\epsffile{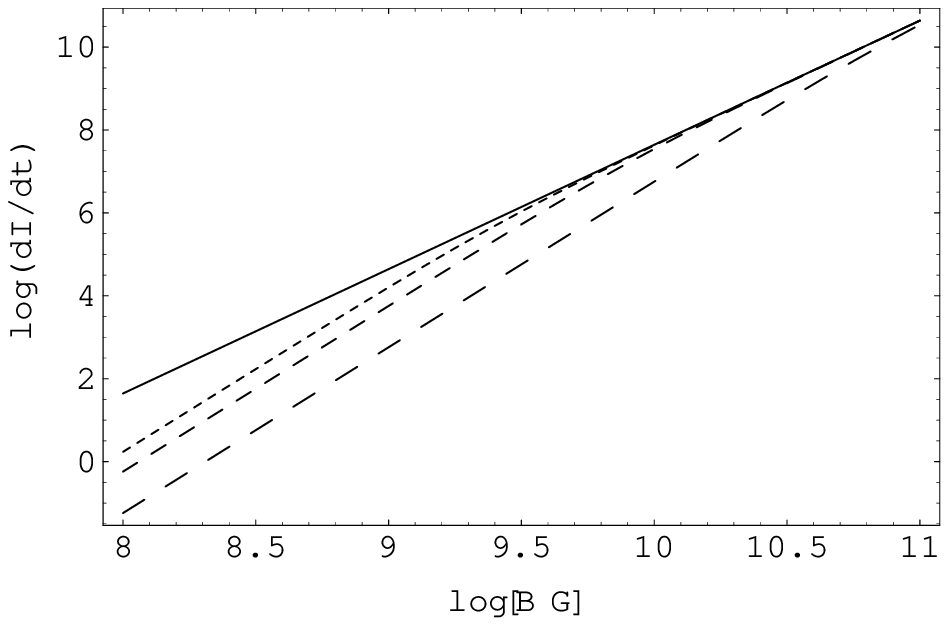}}
\caption{
The rate of energy emission $\frac{dI}{dt}$ during the transition from the Landau level $n=1$ to the
ground state as a function of the magnetic field $B$ (in logarithmic scales)
for the synchrotron mechanism  $R_{c}\rightarrow \infty $ (full curve), $R_c=3\times 10^6 cm$ (dotted curve),
$R_c=10^6 cm$ (dashed curve) and $R_c=10^5 cm$ (long dashed curve) for $\gamma =3\times
10^6$ and $\sin \alpha =10^{-3}$.
}
\label{FIG22}
\end{figure}

\begin{figure}[h]
\epsfxsize=10cm
\centerline{\epsffile{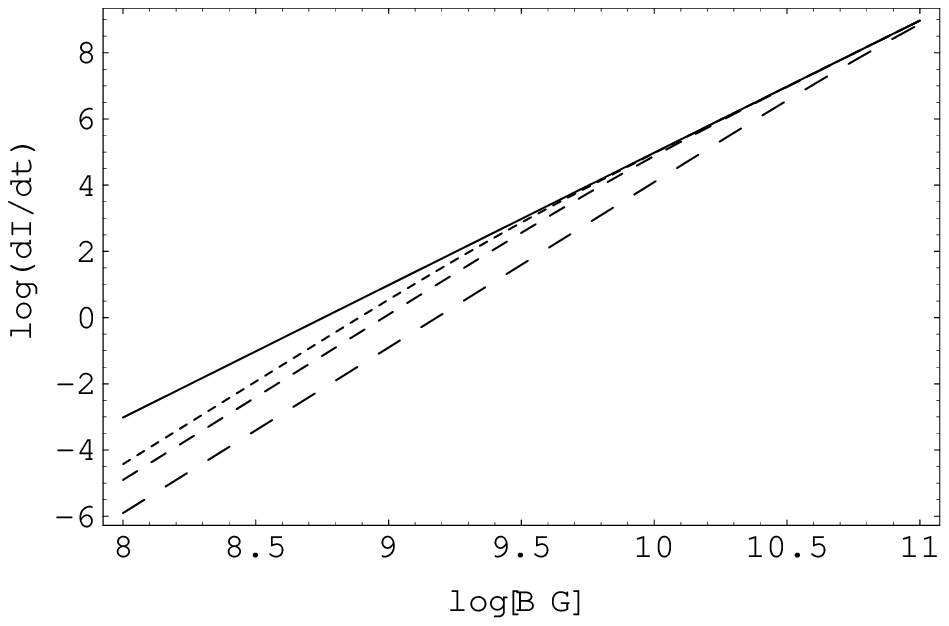}}
\caption{
The rate of energy emission $\frac{dI}{dt}$ during the transition from the Landau level $n=2$ to the
ground state as a function of the magnetic field $B$ (in logarithmic scales)
for synchrotron mechanism $R_{c}\rightarrow \infty $ (full curve), $R_c=3\times 10^6 cm$ (dotted curve),
$R_c=10^6 cm$ (dashed curve) and $R_c=10^5 cm$ (long dashed curve) for $\gamma =3\times
10^6$ and $\sin \alpha =10^{-3}$.
}
\label{FIG23}
\end{figure}

\begin{figure}[h]
\epsfxsize=10cm
\centerline{\epsffile{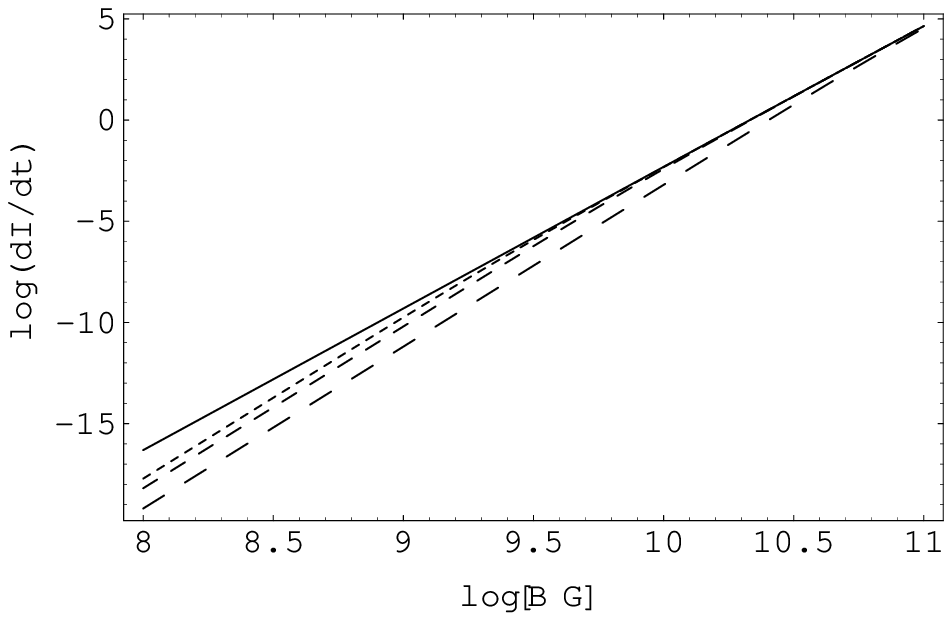}}
\caption{
The rate of energy emission $\frac{dI}{dt}$ during the transition from the Landau level $n=5$ to the
ground state as a function of the magnetic field $B$ (in logarithmic scales) 
for the synchrotron mechanism $R_{c}\rightarrow \infty $ (full curve), $R_c=3\times 10^6 cm$ (dotted curve),
$R_c=10^6 cm$ (dashed curve) and $R_c=10^5 cm$ (long dashed curve) for $\gamma =3\times
10^6$ and $\sin \alpha =10^{-3}$.
}
\label{FIG24}
\end{figure}

\begin{figure}[h]
\epsfxsize=10cm
\centerline{\epsffile{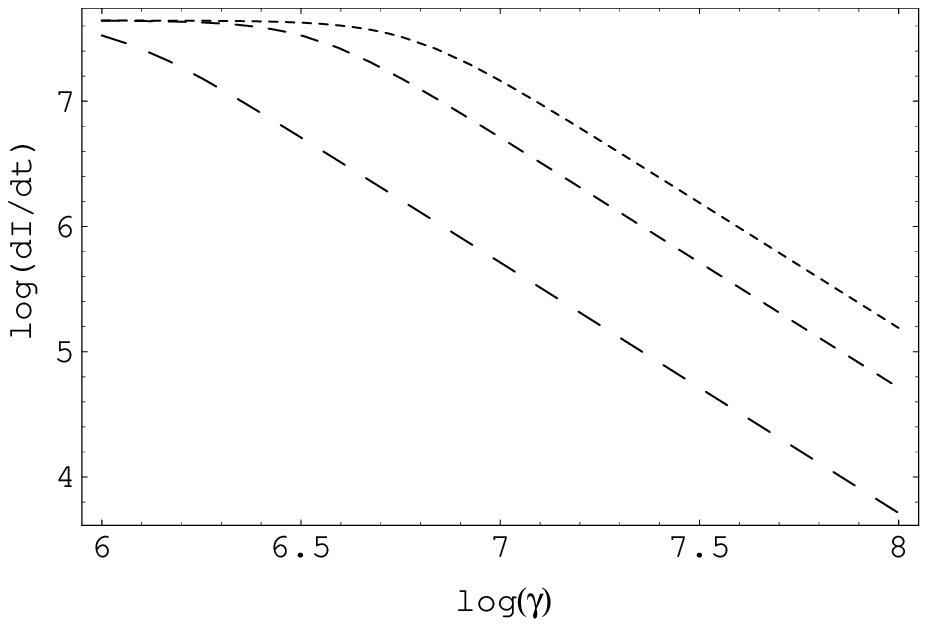}}
\caption{
The rate of energy emission $\frac{dI}{dt}$ during the transition from the Landau level $n=1$ to the
ground state as a function of $\gamma $ (in logarithmic scales) in the generalized
radiation mechanism for $R_c=3\times 10^6 cm$ (dotted curve),
$R_c=10^6 cm$ (dashed curve) and $R_c=10^5 cm$ (long dashed curve) for $B=10^{10} G$
and $\sin \alpha =10^{-3}$.
}
\label{FIG25}
\end{figure}

\begin{figure}[h]
\epsfxsize=10cm
\centerline{\epsffile{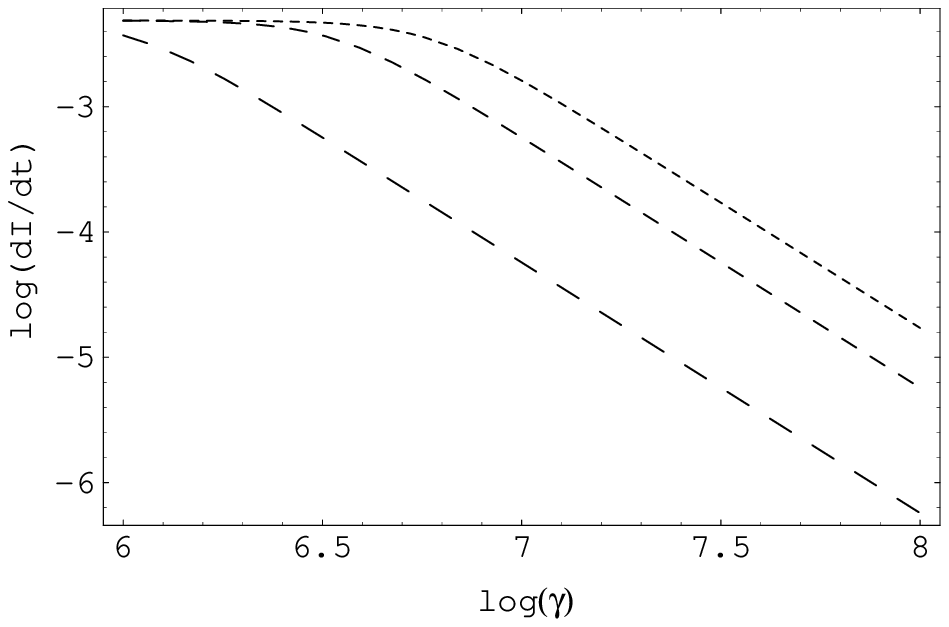}}
\caption{
The rate of energy emission $\frac{dI}{dt}$ during the transition from the Landau level $n=5$ to the
ground state as a function of $\gamma $ (in logarithmic scales)
in the generalized radiation mechanism for $R_c=3\times 10^6 cm$ (dotted curve),
$R_c=10^6 cm$ (dashed curve) and $R_c=10^5 cm$ (long dashed curve) for $B=10^{10} G$
and $\sin \alpha =10^{-3}$.
}
\label{FIG26}
\end{figure}

\begin{figure}[h]
\epsfxsize=10cm
\centerline{\epsffile{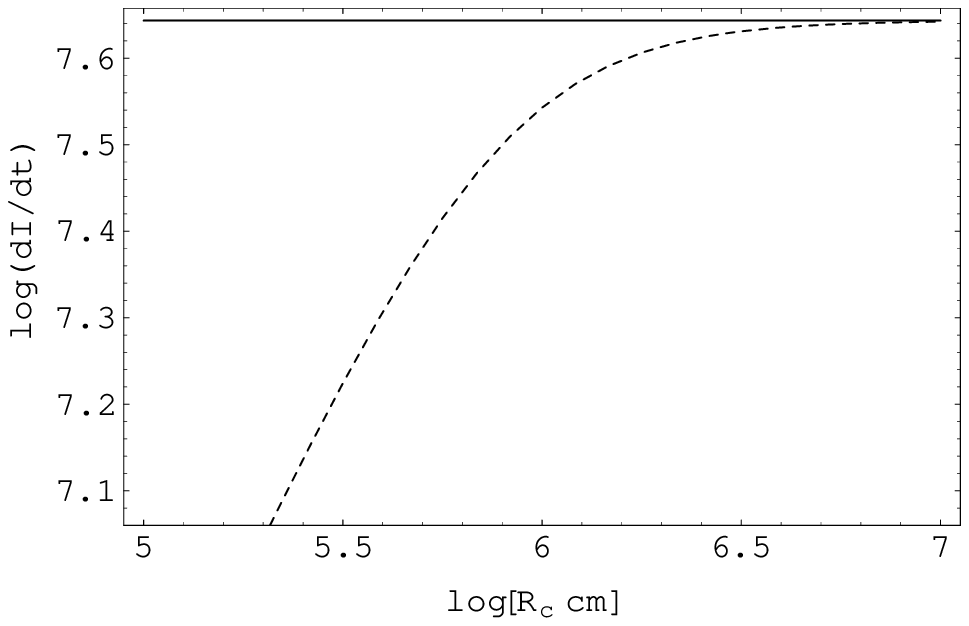}}
\caption{
The rate of energy emission $\frac{dI}{dt}$ during the transition from the Landau level $n=1$ to the
ground state as a function of the radius of curvature $R_c$ (in logarithmic scales) 
for the synchrotron mechanism $R_{c}\rightarrow \infty $(full curve) and generalized radiation mechanism (dotted curve) 
for $B=10^{10} G$, $\gamma =3\times 10^6$ and $\sin \alpha =10^{-3}$. 
}
\label{FIG27}
\end{figure}

\begin{figure}[h]
\epsfxsize=10cm
\centerline{\epsffile{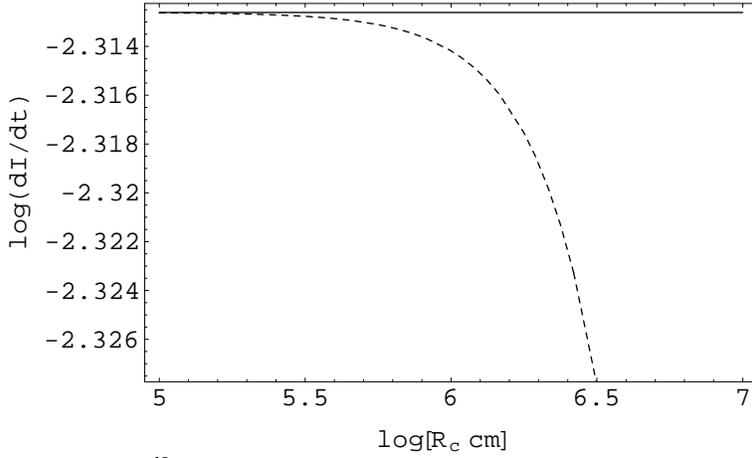}}
\caption{
The rate of energy emission $\frac{dI}{dt}$ during the transition from the Landau level $n=5$ to the
ground state as a function of the radius of curvature $R_c$ (in logarithmic scales) 
for the synchrotron mechanism $R_{c}\rightarrow \infty $(full curve) and generalized radiation mechanism (dotted curve) 
for $B=10^{10} G$, $\gamma =3\times 10^6$ and $\sin \alpha =10^{-3}$. 
}
\label{FIG28}
\end{figure}

The angular distribution of the radiation
emitted during the transition to the ground state follows from $\frac{%
d^{2}I_{n,0}^{(s)}}{dtd\theta }=\hbar \omega \frac{d\Gamma _{n,0}^{(s)}}{%
d\theta }$. Figs. 29-33 present the effect of the gradient drift, described by the radius of curvature of the field, on the
angular distribution of the radiation emitted during transition from low Landau levels $n=1,2,...,5$ to the ground state
$n=0$, for a constant $\gamma $.

\begin{figure}[h]
\epsfxsize=10cm
\centerline{\epsffile{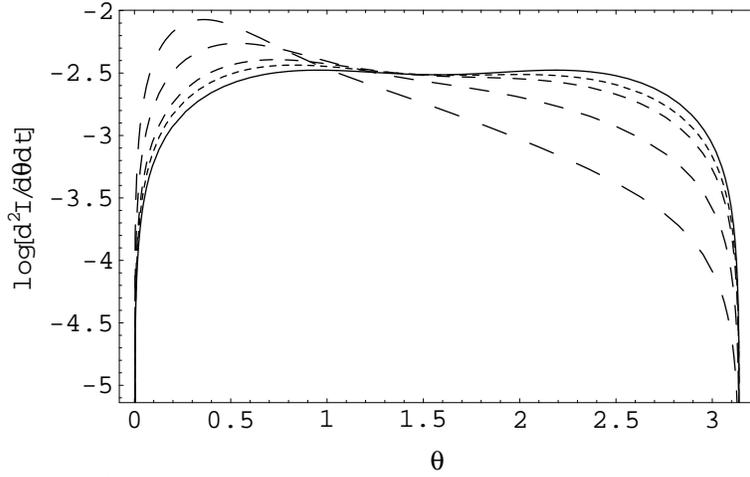}}
\caption{
Angular distribution $\frac{d^{2}I_{n,0}^{(s)}}{dtd\theta }$ (in a logarithmic scale) of the radiation emitted  during the transition from the Landau level $n=1$ to the
ground state for different values of the radius of curvature: 
$R_{c}\rightarrow \infty $ (full curve) (synchrotron mechanism), $R_c=10^7 cm$ (dotted curve), $R_c=5\times 10^6 cm$ (dashed curve),
$R_c=2\times 10^6 cm$ (long dashed curve) and $R_c=10^6 cm$ (ultra-long dashed curve) for
$\gamma =3\times 10^6$, $B=10^{10} G$ and $\sin \alpha =10^{-3}$.
}
\label{FIG29}
\end{figure}

\begin{figure}[h]
\epsfxsize=10cm
\centerline{\epsffile{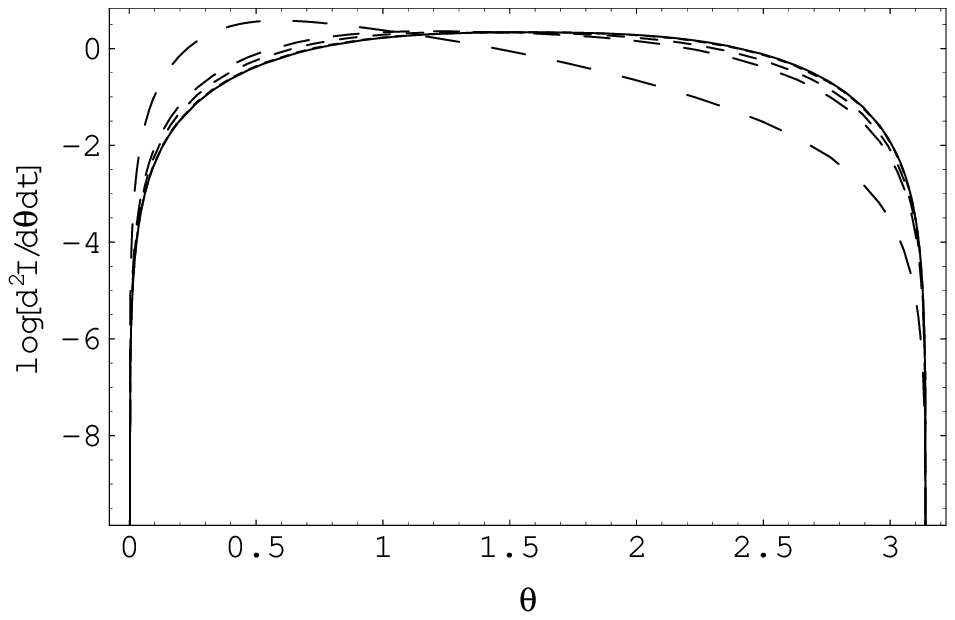}}
\caption{
Angular distribution $\frac{d^{2}I_{n,0}^{(s)}}{dtd\theta }$ (in a logarithmic scale) of the radiation emitted  during the transition from the Landau level $n=2$ to the
ground state for different values of the radius of curvature: 
$R_{c}\rightarrow \infty $ (full curve) (synchrotron mechanism), $R_c=10^7 cm$ (dotted curve), $R_c=5\times 10^6 cm$ (dashed curve),
$R_c=2\times 10^6 cm$ (long dashed curve) and $R_c=10^6 cm$ (ultra-long dashed curve) for
$\gamma =3\times 10^6$, $B=10^{10} G$ and $\sin \alpha =10^{-3}$.
}
\label{FIG30}
\end{figure}

\begin{figure}[h]
\epsfxsize=10cm
\centerline{\epsffile{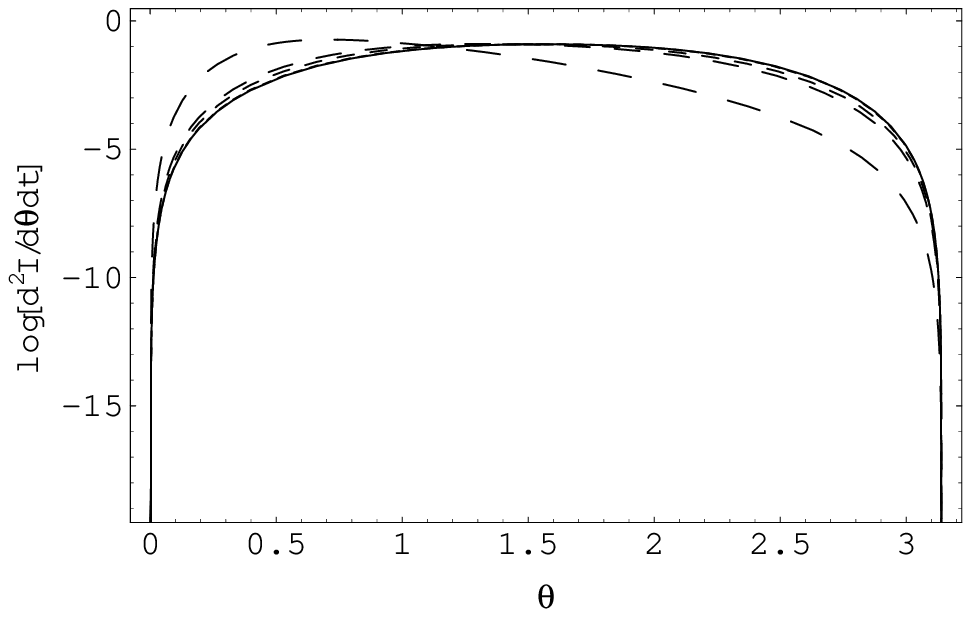}}
\caption{
Angular distribution $\frac{d^{2}I_{n,0}^{(s)}}{dtd\theta }$ (in a logarithmic scale) of the radiation emitted  during the transition from the Landau level $n=3$ to the
ground state for different values of the radius of curvature: 
$R_{c}\rightarrow \infty $ (full curve) (synchrotron mechanism), $R_c=10^7 cm$ (dotted curve), $R_c=5\times 10^6 cm$ (dashed curve),
$R_c=2\times 10^6 cm$ (long dashed curve) and $R_c=10^6 cm$ (ultra-long dashed curve) for
$\gamma =3\times 10^6$, $B=10^{10} G$ and $\sin \alpha =10^{-3}$.
}
\label{FIG31}
\end{figure}

\begin{figure}[h]
\epsfxsize=10cm
\centerline{\epsffile{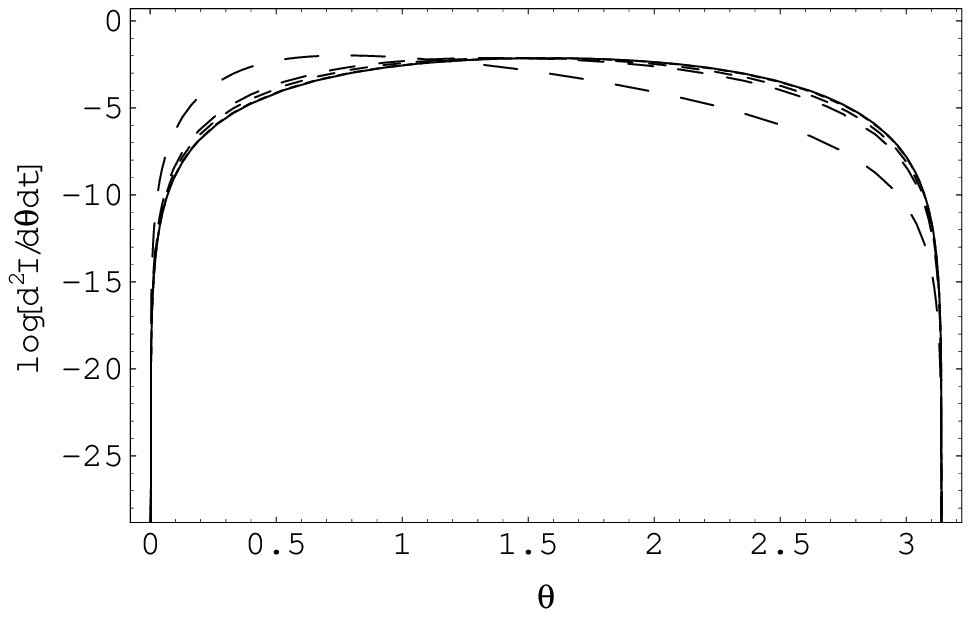}}
\caption{
Angular distribution $\frac{d^{2}I_{n,0}^{(s)}}{dtd\theta }$ (in a logarithmic scale) of the radiation emitted  during the transition from the Landau level $n=4$ to the
ground state for different values of the radius of curvature: 
$R_{c}\rightarrow \infty $ (full curve) (synchrotron mechanism), $R_c=10^7 cm$ (dotted curve), $R_c=5\times 10^6 cm$ (dashed curve),
$R_c=2\times 10^6 cm$ (long dashed curve) and $R_c=10^6 cm$ (ultra-long dashed curve) for
$\gamma =3\times 10^6$, $B=10^{10} G$ and $\sin \alpha =10^{-3}$.
}
\label{FIG32}
\end{figure}

\begin{figure}[h]
\epsfxsize=10cm
\centerline{\epsffile{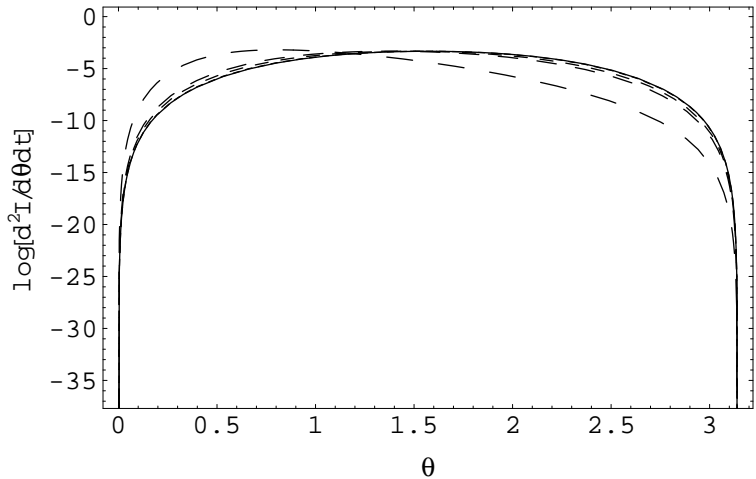}}
\caption{
Angular distribution $\frac{d^{2}I_{n,0}^{(s)}}{dtd\theta }$ (in a logarithmic scale) of the radiation emitted  during the transition from the Landau level $n=5$ to the
ground state for different values of the radius of curvature: 
$R_{c}\rightarrow \infty $ (full curve) (synchrotron mechanism), $R_c=10^7 cm$ (dotted curve), $R_c=5\times 10^6 cm$ (dashed curve),
$R_c=2\times 10^6 cm$ (long dashed curve) and $R_c=10^6 cm$ (ultra-long dashed curve) for
$\gamma =3\times 10^6$, $B=10^{10} G$ and $\sin \alpha =10^{-3}$.
}
\label{FIG33}
\end{figure}

\section{Discussions and final remarks}

In the present paper we have proposed, based on simple physical considerations,
a unified formalism to describe the radiation emitted by an ultra-relativistic electron
moving in an inhomogeneous magnetic field in both classical and quantum domains.
In the classical domain we have derived, by using Schwinger's approach, a general
representation of the radiation patterns of a relativistic electron in an arbitrary motion.
The classical radiation spectrum and the characteristic frequency can be generally expressed
as a function of the absolute value of the acceleration of the particle.
These general expressions have been used to obtain the distribution over the frequencies
of the radiation of the electron in an inhomogeneous magnetic field.

This model combines in a single expression the well-known synchrotron and curvature mechanisms and
adds a new correction term involving the magnetic gradient drift of the electron due to the inhomogeneous
character of the magnetic field. Although from numerical point of view our results are similar  to those obtained
in the previous investigations (Zhang \& Cheng 1995; Cheng \& Zhang 1996; Zhang et al. 2000), our formulae
are dealing with a more realistic magnetic field. Of course this monopole-type field
can only be considered as an approximation of the realistic (dipole-type) magnetic field
of pulsars. The differences in the magnetic field model
lead to the differences in the spectral distribution of the radiation between the
present model and the Cheng \& Zhang (1996) model.

All the exact results derived by using Schwinger's method (Schwinger et al. 1976) can be also obtained by
using the following prescription. From a physical point of view the curvature and gradient terms can be formally described
as an electric field acting on the electron. This electric field and the magnetic field
can be combined into a single effective electromagnetic field which is then used as a substitute
for the (constant) magnetic field in the well-known expressions of the synchrotron radiation.
This physical approach to the general radiation mechanism is consistent with the
rigurous classical description, leading to the same physical results, thus giving an exact description of the
radiation process from regions in which the magnetic field intensity and the radius of curvature
of the electron's trajectory are slowly varying.

On the other hand, due to its simplicity,
the present approach based on the substitution of the constant magnetic field with the
effective field can be easily extended to the quantum domain of the radiation spectrum,
leading to the possibility of a complete description of the quantum recoil due to the
emission of a photon. The main advantage of the present approach relies in its mathematical
simplicity, allowing the derivation of quite simple analytical expressions for all the physical
characteristics of the radiation of an electron in a strong magnetic field, when quantum
effects become important. When curvature effects cannot
be neglected our formulae can take care of both regimes smoothly. 

The application in the quantum domain of the formalism developed in the present paper
is restricted to a region of strengths of magnetic
fields so that $B\le B_{0}$, that is to values of the magnetic field smaller than $4\times 10^{13}G$.
Recent observations show that magnetic fields with strengths much larger than this value
can also exist on the surface of compact objects.
We must also mention that in the discussion of the quantum effects we have
sistematically neglected the effects due to the spin of the electron.

\section*{Acknowledgement}

This work is partially supported by a RGC grant of the Hong Kong government. T. H. is
supported by a postdoctoral fellowship of the University of Hong Kong. We thank J. L. Zhang, Y. F. Yuan and
D. C. Wei for useful discussions and Anisia Tang for help in the preparation
of the manuscript.

\section*{Appendix}

In order to calculate the integral
\begin{equation}
I=\int_{0}^{\infty }\left( 1+2x^{2}\right) \left\{ \sin \left[ \frac{3}{2}%
\xi \left( x+\frac{1}{3}x^{3}\right) \right] -\sin bx\right\} \frac{dx}{x},
\end{equation}
with $\xi >0$ and $b>0$ constants, we use first the well-known result $%
\int_{0}^{\infty }\frac{\sin bx}{x}dx=\frac{\pi }{2}$. Hence the integral
becomes
\begin{equation}
I=\int_{0}^{\infty }\left( 1+2x^{2}\right) \sin \left[ \frac{3}{2}\xi \left(
x+\frac{1}{3}x^{3}\right) \right] \frac{dx}{x}-\frac{\pi }{2}.
\end{equation}

With the use of the Airy integral (Abramowitz \& Stegun 1972),
\begin{equation}
\frac{1}{\sqrt{3}}K_{1/3}\left( \xi \right) =\int_{0}^{\infty }\cos \left[ 
\frac{3}{2}\xi \left( x+\frac{1}{3}x^{3}\right) \right] dx,
\end{equation}
we obtain  
\begin{equation}
\frac{1}{\sqrt{3}}K_{1/3}\left( \xi \right) =\frac{d}{d\xi }\int_{0}^{\infty
}\sin \left[ \frac{3}{2}\xi \left( x+\frac{1}{3}x^{3}\right) \right] \frac{dx%
}{x}.
\end{equation}

Hence,
\begin{eqnarray}
\frac{1}{\sqrt{3}}\int_{\xi }^{\infty }K_{1/3}\left( \eta \right) d\eta 
&=&\lim_{L\rightarrow \infty }\int_{0}^{\infty }\sin \left[ L\left( x+\frac{1%
}{3}x^{3}\right) \right] \frac{dx}{x}-\int_{0}^{\infty }\sin \left[ \frac{3}{%
2}\xi \left( x+\frac{1}{3}x^{3}\right) \right]  \nonumber\\
&=&\frac{\pi }{2}-\int_{0}^{\infty }\sin \left[ \frac{3}{2}\xi \left( x+%
\frac{1}{3}x^{3}\right) \right] \frac{dx}{x}.
\end{eqnarray}

On the other hand from the general relation $\left( \frac{d}{d\xi }+\frac{1}{%
3\xi }\right) K_{1/3}\left( \xi \right) =-K_{2/3}\left( \xi \right) $ (Abramowitz \& Stegun 1972) we
obtain $\frac{1}{\sqrt{3}}K_{2/3}\left( \xi \right) =\int_{0}^{\infty }x\sin %
\left[ \frac{3}{2}\xi \left( x+\frac{1}{3}x^{3}\right) \right] dx$.
Therefore we obtain
\begin{equation}
I=\frac{1}{\sqrt{3}}\left( 2K_{2/3}\left( \xi \right) -\int_{\xi }^{\infty
}K_{1/3}\left( \eta \right) d\eta \right) =\frac{1}{\sqrt{3}}\int_{\xi
}^{\infty }K_{5/3}\left( \eta \right) d\eta ,
\end{equation}
where we have also used the reccurence relation $2\frac{d}{d\xi }%
K_{2/3}\left( \xi \right) +K_{1/3}\left( \xi \right) =-K_{5/3}\left( \xi
\right) $ (Abramowitz \& Stegun 1972). 

\end{document}